\PassOptionsToPackage{prologue,dvipsnames}{xcolor}
\documentclass{article}
\usepackage{lmodern}
\usepackage{amsmath,amssymb,amsthm}

\usepackage{booktabs}
\usepackage{threeparttable}
\usepackage{multirow}
\usepackage{graphicx}
\usepackage[font=small]{caption}
\usepackage{subcaption}
\usepackage{adjustbox}
\usepackage{float}
\usepackage{rotating}
\usepackage{placeins}
\usepackage{tabularx}
\usepackage{arydshln}
\usepackage{array}
\usepackage{etoolbox}
\usepackage{pifont}
\usepackage{comment}

\usepackage[a-1b]{pdfx}
\hypersetup{hidelinks}
\usepackage{xr-hyper}
\usepackage{pdflscape}
\usepackage[authoryear,round]{natbib}

\usepackage{titlesec}
\titleformat{\section}{\normalfont\large\bfseries}{\thesection}{1em}{}
\titleformat{\subsection}{\normalfont\normalsize\bfseries}{\thesubsection}{1em}{}
\titleformat{\subsubsection}{\normalfont\small\bfseries}{\thesubsubsection}{1em}{}
\usepackage{arxiv}

\title{Landmarking with Latent Class Mixed Models for Dynamic Prediction of Time-to-event Data with Heterogeneous Biomarker Trajectories}

\author{
 Víctor Velasco-Pardo\thanks{To whom correspondence should be addressed.}\\
  Institute of Genetics and Cancer, \\
  University of Edinburgh, \\
  Edinburgh, UK \\
  \texttt{vvelasco@ed.ac.uk} \\
   \And
 Nathan Constantine-Cooke \\
  Institute of Genetics and Cancer, \\
  University of Edinburgh, \\
  Edinburgh, UK \\
  \texttt{nathan.constantine-cooke@ed.ac.uk} \\
  \And
 Charlie W. Lees \\
  Institute of Genetics and Cancer, \\
  University of Edinburgh, \\
  Edinburgh, UK \\
  \texttt{charlie.lees@ed.ac.uk} \\
  \And
 Catalina A. Vallejos\footnotemark[1] \\
  Institute of Genetics and Cancer, \\
  University of Edinburgh, \\
  Edinburgh, UK \\
\texttt{Catalina.vallejos@ed.ac.uk} \\
}

\begin{document}

\maketitle


\begin{abstract}
{
The increasing ability to securely access electronic health records (EHR) has created unprecedented opportunities to monitor the health trajectories of large heterogeneous patient populations throughout their lifetime. Repeated measurements of time-varying covariates (such as biomarkers measured via routine blood tests) can inform dynamic risk prediction of time-to-event outcomes, updating risk estimates as new information becomes available. 
Existing dynamic risk prediction approaches often assume homogeneous longitudinal trajectories across individuals. This assumption is not met when 
there is heterogeneity driven by latent subgroup structure (e.g.~due to unobserved confounders), as is often the case with real-world biomedical data. At present, accounting for such heterogeneity is only available in joint latent class models for longitudinal and time-to-event data, but they are computationally intensive, often prohibitively so for large-scale data, such as those present in EHR settings. To address these challenges, we propose a novel landmarking approach that integrates latent class mixed models (LCMMs) to capture latent heterogeneity in longitudinal trajectories. 
Our method is implemented in a modular R package, \texttt{landmaRk}, which is available on CRAN and allows users to flexibly specify the components of a landmarking analysis, beyond our proposed approach. Through simulation studies, we demonstrate improvements in prediction performance in the presence of latent heterogeneity compared to traditional landmarking strategies, while remaining computationally efficient for large datasets. We also provide a proof-of-concept illustration using real data.}
\end{abstract}


\section{Introduction}
\label{sec::intro}

In time-to-event analyses, it is common to observe repeated measurements of time-varying covariates that are related to the event of interest. Among others, this can include longitudinal trajectories for biomarkers used to monitor disease progression, which are widely available in electronic health records \citep[EHR;][]{gao_pipeline_2025}. 
An important aim in this context is to exploit the process underpinning those longitudinal measurements to make accurate prognostic predictions. This can be thought of as a dynamic risk prediction problem, where the goal is to estimate the probability of being event-free up to a given time ($s+w$) conditioned on still being at risk at a certain point after baseline ($s>0$) and the trajectory of observed longitudinal measurements up until $s$ \citep{schumacher_dynamic_2020}.   

There are two popular dynamic risk prediction approaches in the biostatistics literature: joint models for longitudinal and survival (JMLS) data and landmarking \citep{rizopoulos_dynamic_2017}. JMLS specify a joint likelihood to define longitudinal and time-to-event sub-models, often using a shared random effects specification \citep[e.g.][]{wulfsohn_joint_1997}. Instead, landmarking is a two-step approach where separate time-to-event models are defined for the subset of individuals still at risk at a pre-specified sequence of landmark times \citep[superlandmarking extends this to borrow information across landmarks,][]{de_swart_comparative_2025}. At each landmark time, predictions are conditioned on the longitudinal measurements observed thus far, typically using a last observation carried forward (LOCF) approach \citep{van_houwelingen_dynamic_2007} or based on predictions from a linear mixed effects (LME) model \citep{paige_landmark_2018}.

JMLS have been shown to outperform landmarking when the model is correctly specified \citep{ferrer_individual_2019}. However, the use of JMLS in the context of EHRs can be challenging, as longitudinal information is available for large heterogeneous cohorts, often over varying periods of time \citep{paige_landmark_2018}. JMLS can be computationally expensive, failing to scale well for such datasets  \citep{rizopoulos_dynamic_2017}. Moreover, JMLS require the definition of a common baseline ($t = 0$) from which longitudinal measurements are recorded. This does not allow for the varying levels of longitudinal follow-up present in EHRs, e.g. due to the in/outflow of individuals over time \citep{paige_landmark_2018}.  
Landmarking provides a flexible approach that bypasses these limitations; for example, by allowing analysts to choose how much of the longitudinal history up to the landmark time is incorporated into the model.

JMLS and landmarking 
typically assume the process underlying longitudinal measurements is largely homogeneous across individuals, with inter-individual heterogeneity captured by observed covariates (e.g.~fixed effects in an LME) or through random effects with a thin-tailed distribution (e.g.~normal). 
This assumption is not met in 
contexts where there is latent population structure related to distinct groups of individuals who share similar longitudinal trajectories, and where such heterogeneity is also reflected into differences in risk profiles.
Examples include biomarkers of inflammation for patients with inflammatory bowel disease 
\citep{constantine-cooke_longitudinal_2023,constantine-cooke_large-scale_2025}, cognitive tests and biomarkers used to monitor Alzheimer's disease progression \citep{howlett_disease_2021}, and biomarkers used in the context of diabetic kidney disease 
\citep{jiang_progression_2019}. 
Joint latent class models (JLCM) \citep{lin_latent_2000,lin_latent_2002} are an exception but, similar to JMLS, their use in the context of EHRs is challenging due to computational costs and the requirement of following individuals from a common baseline time ($t=0$). 
As such, there is an unmet need for scalable dynamic risk prediction approaches which account for latent heterogeneity in longitudinal trajectories.

Our contributions in this article are twofold. First, we present a novel heterogeneity-aware dynamic risk prediction strategy that integrates latent class mixed models \citep[LCMM,][]{mcculloch_discovering_2002} into landmarking, leveraging the implementation of \cite{proust-lima_estimation_2017}. Secondly, we present a flexible \texttt{R} package for landmarking analysis, \texttt{landmaRk}, which is freely available on CRAN and at \url{https://github.com/VallejosGroup/landmaRk}. This formalises landmarking as a modular pipeline of interchangeable components, allowing flexible extensions of the landmarking framework, 
with user-defined arbitrary specifications for the longitudinal and survival components. 
Finally, using \texttt{landmaRk}, we illustrate the use of our proposed heterogeneity-aware landmarking approach, and evaluate its performance, with both simulated and real data. 



\section{Model specification}
\label{sec:model-specification}

\subsection{Notation}

We assume a dataset $\{T_i, D_i, \mathcal{Y}_i, \mathbf{x}_i, \mathbf{w}_i\}_{i=1}^N$, where $N$ denotes the number of individuals, $T_i$ is the event or censoring time 
for individual $i$ and $D_i$ is a censoring indicator ($1$ if an event occurred before censoring for individual $i$; $0$ otherwise). For each individual $i$, $\mathbf{x}_i$ and $\mathbf{w}_i$ denote two vectors of static covariates associated to the longitudinal and survival outcomes, respectively (the covariates in $\mathbf{x}_i$ and $\mathbf{w}_i$ may overlap). Moreover, $\mathcal{Y}_i = (y_{i1},\dots,y_{in_i})^\top$ denotes a vector of $n_i$ repeated measurements for a time-varying covariate. We assume $y_{ij} = y_{i}(t_{ij})$ is a noisy measurement of a latent covariate process $m_i(t)$, recorded at time $t_{ij}$. Note that the time points where longitudinal measurements are recorded, and the number of measurements $n_i$ may differ across individuals.

\subsection{Landmarking} \label{subsec:landmarking}

\begin{figure}[H]
\centering\includegraphics[width=\textwidth]{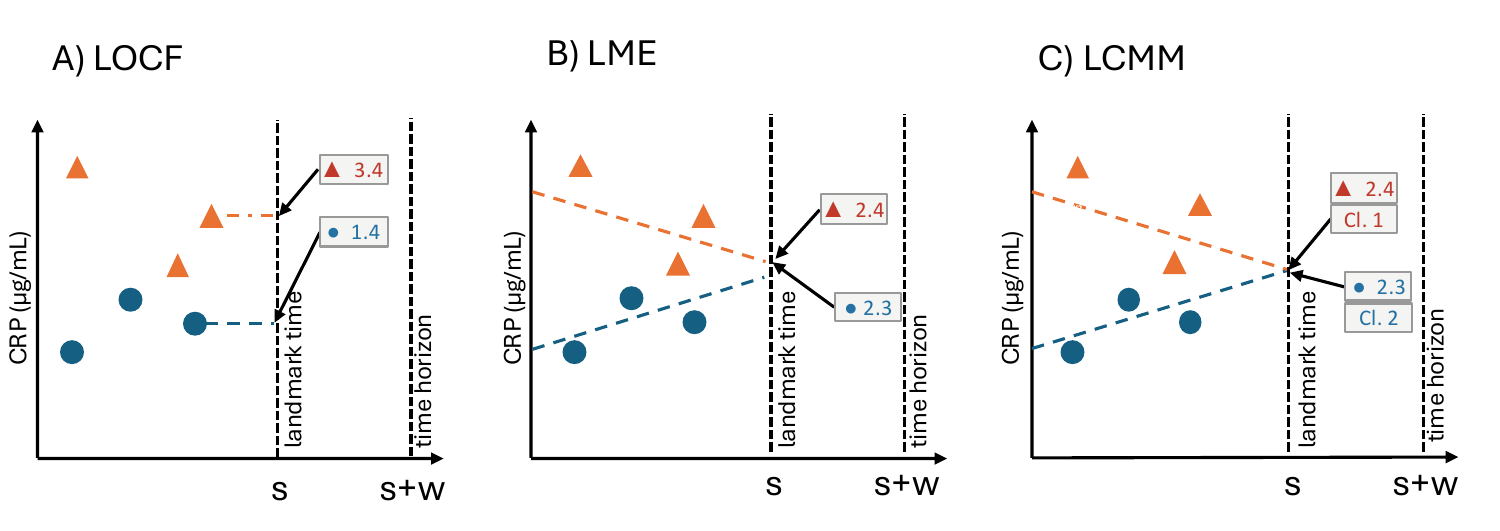}
        \caption{Schematic illustration of landmarking with different strategies (LOCF, LME, and LCMM) to summarise the longitudinal trajectory $\mathcal{Y}_i(s)$ at each landmark time $s$. 
        The derived summary, $\hat{y}_i(s)$ is then used as an input to a survival submodel, which is used for prediction over the prediction window $[s, s+w)$. (A) LOCF uses the last observation as a summary. (B) LME uses a prediction for the latent trajectory, taking into account measurement error but without accounting for latent heterogeneity. (C) Our proposed heterogeneity-aware approached based on LCMM uses a prediction for the latent trajectory, accounting for latent heterogeneity arising due to subpopulation structure, and optionally the predicted cluster label. LME and LCMM can also account for heterogeneity that is explained by  $\mathbf{x}_i$ (observed covariates) via fixed effects. Blue circles and orange triangles represent repeated measurements of a longitudinal covariate for two different individuals prior to the landmark; dashed lines indicate the inferred trajectories that are used to derive summaries $\hat{y}_i(s)$ at the landmark time $s$. Vertical dashed lines denote the prediction window.} 
        \label{fig:diagram}
\end{figure}

Landmarking consists of fitting survival models in a sequential manner, iterating over a set of pre-specified landmark times \citep{van_houwelingen_dynamic_2011}. At each landmark $s$, the first task is to identify the risk set $\mathcal{R}_s$, i.e.~the set of individuals that have not experienced the event (or are censored) prior to time $s$. The second task is to summarise the trajectory of the time-varying covariates observed before the landmark time, denoted as $\mathcal{Y}_i(s)$, where $\mathcal{Y}_i(s) = \{ (y_{i1},\dots,y_{il})^\top \text{ such that } t_{il} \leq s \}$. 
Such summary is then used as an input for landmark-specific survival models. The simplest and most common approach is ``Last Observation Carried Forward'' (LOCF) \citep[Figure \ref{fig:diagram}A;][]{van_houwelingen_dynamic_2007}. Alternatively, one can fit an 
LME, 
and use the best linear unbiased predictor 
for $Y_i(s)$ as a summary \citep[Figure \ref{fig:diagram}B;][]{barrett_dynamic_2017}. This combines a population-level prediction (fixed effects) with the predicted random effects (based on the subject-specific longitudinal measurements observed up to $s$). The first approach fails to account for measurement error, whilst the latter relies on the assumption of homogeneous random effects implicit in the LME. 


Subsequently, a conditional landmark-specific survival regression model is fitted, with the aforementioned summary of $\mathcal{Y}_i(s)$ (which we denote as $\hat{y}_i(s)$) as an input. Dynamic risk prediction is then performed by estimating \begin{equation} \label{eq:dyn_prob}
    \pi_i(s + w \mid s) = \text{Pr}(T_i > s + w \mid T_i \ge s, \hat{y}_i(s), \mathbf{w}_i), \quad w > 0,
\end{equation} for the set of individuals in the risk set $\mathcal{R}_s$. Estimation of (\ref{eq:dyn_prob}) is typically done using a landmark-specific Cox Proportional Hazards \citep[PH; ][]{cox_regression_1972} model (but any survival model could be used as an alternative), defined as 
\begin{equation}
    h_i(t \mid \mathbf{w}_i, \hat{y}_i(s), s) = h_{0}(t \mid s) \exp\!\left\{ \boldsymbol{\gamma}_s^\top \mathbf{w}_i + \alpha_s \hat{y}_i(s) \right\}, \quad t > s,
\end{equation}
where $\boldsymbol{\gamma}_s$ and $\alpha_s$ are landmark-specific parameters, and $h_0(t \mid s)$ is the baseline hazard from the landmark time $s$. To fit this model, survival times are often censored at $s+w$ to relax the implicit assumption that time-varying covariates remain constant over $[s, s+w)$ \citep[the so-called problem of ``ageing covariates'';][]{van_houwelingen_dynamic_2011}. Alternatively, 
one can fit a single survival model to a stacked dataset in which individuals appear as many times as landmark times they have survived. This approach, known as superlandmarking \citep{de_swart_comparative_2025}, borrows strength across landmark times but inference can be challenging for large cohorts \citep{wang_fitting_2024}. 


\subsection{Heterogeneity-aware landmarking} \label{subsec:heterogeneity}

LCMM is a powerful framework which has been used to describe latent population heterogeneity across several disease contexts \citep[e.g.][]{jiang_progression_2019,howlett_disease_2021,constantine-cooke_longitudinal_2023,constantine-cooke_large-scale_2025}. Our goal is to embrace this heterogeneity, 
borrowing information across individuals that share similar longitudinal trajectories and risk profiles, when such heterogeneity cannot be explained by known covariates. 
We hypothesise that leveraging LCMM to explicitly account for latent population heterogeneity within a landmarking context will enable richer information to be extracted from longitudinal trajectories, leading to more accurate estimates of risk. 


We consider LCMM as implemented in the \texttt{lcmm} R package \citep{proust-lima_estimation_2017}. Full details are shown in Supplementary Section \ref{sec:lcmm-logistic-regression}. In brief, 
LCMM has two components, cluster membership and longitudinal submodels, each of which may use a subset of the covariates $\textbf{x}_i$ as an input. The LCMM framework jointly estimates the parameters associated to each sub-model. 
We define the LCMM structure in a landmark-specific manner; that is, for a landmark time $s$, the model only considers information observed prior to $s$ only for individuals in $\mathcal{R}_s$, i.e.~$\left\{\mathcal{Y}_i(s); i \in \mathcal{R}_s\right\}$. This allows the number of underlying clusters and the associated model parameters to vary across landmarks. For example, whilst a single cluster may be chosen for early landmarks (where less information is available), a more complex latent population structure may arise over time. For ease of notation, hereafter we avoid indexing LCMM parameters by $s$.

Cluster allocation indicators for each individual $i$, 
$c_i$, are modelled using a multinomial logistic regression for $\text{Pr}(c_i=g \mid \mathbf{x}_{i})$, $g = 1, \ldots, G$. 
A conditional, cluster-specific longitudinal model is then defined, 
\begin{equation} \label{eq:cluster_lme}
    y_{i}(t_{ij})  \big|_{c_i = g} = m_i(t_{ij} \mid \mathbf{x}_{i}, c_i = g) + \epsilon_{ij}, \quad \epsilon_{ij} \sim \mathcal{N}(0, \sigma^2) \quad i \in \mathcal{R}_s, j \in \{1,\ldots,n_i\}
\end{equation} The latter can be interpreted as a cluster-specific LME, 
where $m_i(t \mid \mathbf{x}_{i}, c_i = g)$ 
defines a latent cluster-specific trajectory. The latter is further decomposed into terms related to fixed 
and random effects capturing
overall cluster-specific patterns 
and inter-individual variation within each cluster, respectively. Within this framework, non-linear longitudinal trajectories can be accommodated; for example, via natural cubic splines or Gaussian radial basis functions. 



For each individual $i$, LCMM provides 
a probabilistic prediction for cluster membership, conditional on their observed longitudinal measurements $\mathcal{Y}_i(s)$. This is derived from Bayes' rule as \begin{equation}
    \pi_{ig} = \text{Pr}(c_i = g \mid \mathbf{x}_i, \mathcal{Y}_i(s)) = \dfrac{\phi_{ig}(\mathcal{Y}_i(s) \mid \mathbf{x}_i, c_i = g) \text{Pr}(c_i = g \mid \mathbf{x}_i)}{\text{Pr}( \mathcal{Y}_i(s) \mid \mathbf{x}_i)}, \quad g=1,\dots,G,
\end{equation}
where $\phi_{ig}(\cdot \mid \mathbf{x}_i, c_i = g)$ is the marginal density for $\mathcal{Y}_i(s)$ after integrating out the cluster-specific random effects and $\text{Pr}( \mathcal{Y}_i(s) \mid \mathbf{x}_i)$ is the marginal likelihood of of the observed longitudinal measurements for individual $i$, after integrating out the cluster labels. Estimates $\hat{\pi}_{ig}$ are calculated based on maximum likelihood estimates for all model parameters (these are referred to as posterior probabilities in the output from \texttt{lcmm}). By default, each individual is allocated to the cluster with the highest predicted probability, $\hat{c}_i = \arg\!\max_i\{\hat{\pi}_{ig}\}$. It is also possible to obtain cluster-specific predictions for the longitudinal process itself, which we here denote as $\hat{y}_i(s | \mathbf{x}_i, c_i = g)$. We consider individual-specific predictions obtained using empirical Bayes estimates for the random effects \citep{proust-lima_estimation_2017}.




We use the LCMM output 
to 
summarise of the longitudinal trajectory up to the landmark time $s$, as $\hat{y}_i(s)$. One option is to use predictions based on the predicted cluster, that is, $\hat{y}_i(s) = \hat{y}_i(s \mid \mathbf{x}_i, c_i = \hat{c}_i)$. However, this would ignore uncertainty around cluster allocations which can be substantial, particularly for individuals with fewer observations or shorter longitudinal follow-ups. Instead, we take full account of the uncertainty around predicted cluster membership by averaging cluster-specific trajectories according to individual-specific predicted probabilities, 
that is 
\begin{equation}
    \hat{y}_i(s) = \sum\limits_{g=1}^G \hat{\pi}_{ig} \hat{y}_i(s \mid \mathbf{x}_i, c_i = g)\label{eq:summary-lcmm}.
\end{equation}

We also consider adding the predicted cluster label, $\hat{c}_i$, as an input for the landmark-specific survival models. Indeed, our simulations (Section \ref{sec:simulation}) show that including the predicted cluster label can have a substantial effect, 
as the empirical Bayes predictions $\hat{y}_i(s)$ are, in our experience, largely similar across LME and LCMMs. The predicted cluster labels provide additional information about the trajectory shape, which potentially carries prognostic information beyond the predicted value at the landmark time. In principle, information about the trajectory shape could be obtained from the LME by passing the predicted random effects directly to the survival submodel. However, this may require a large number of random effects to capture non-linear trajectory patterns (e.g.~via natural cubic splines), whereas the cluster label provides a parsimonious and interpretable summary of trajectory shape.


Fitting landmark-specific LCMMs, ensures that the longitudinal model is not biased by informative truncation of the measurement process by the event of interest, as individuals who have already experienced the event do not contribute to the trajectory estimates \citep{barrett_dynamic_2017}. However, in practice, the optimiser implemented in the LCMM R package may fail to converge when few repeated measurements are available (early landmarks) or where the risk set $\mathcal{R}_s$ is small (late landmarks). In those cases, one can fit a single LCMM to all available longitudinal measurements, regardless of when events occur. This leads to further computational gains, but does not condition on survival to $s$ and may therefore introduce the aforementioned bias.

\section{The \texttt{landmaRk} package} \label{sec:landmaRk}
A key strength of the landmarking approach to dynamic risk prediction is its flexibility: in principle, any longitudinal model and any survival model can be used, respectively, to summarise the longitudinal trajectories and to perform risk prediction. This was demonstrated e.g., by  \cite{tanner_dynamic_2021} using a machine learning ensemble framework. However, to the best of our knowledge, there is no ready-to-use 
software 
that takes full advantage of this flexibility.

\begin{figure}[t]
\centering\includegraphics[width=\textwidth]{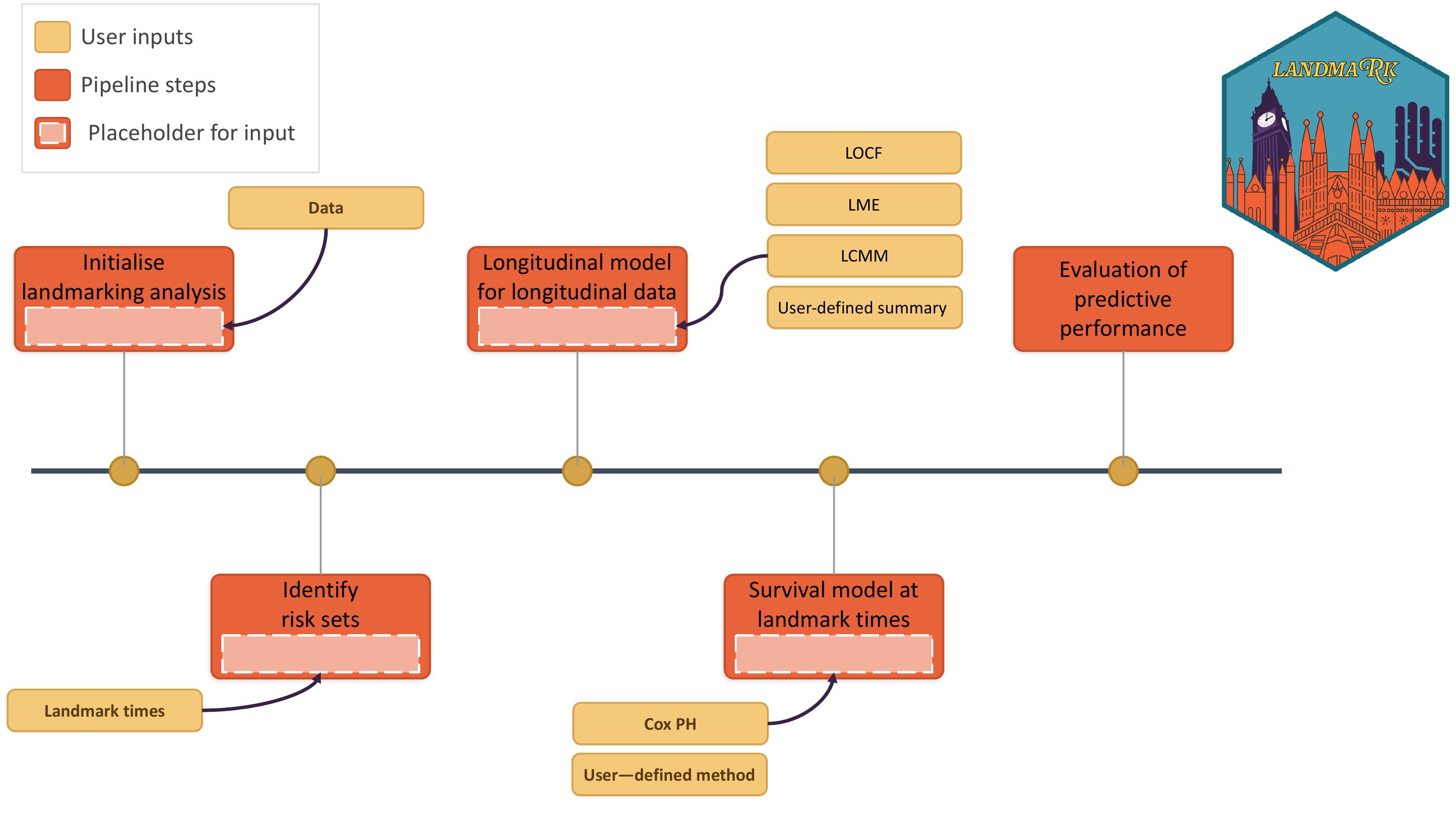}
        \caption{The typical \texttt{landmaRk} pipeline. The \texttt{landmaRk} R package allows analyst to implement landmarking analyses via the following steps: 1) Initialise the landmarking analysis, 2) specify the risk sets (for a pre-specified set of landmark times, 3) fit the longitudinal submodel, 4) fit a survival submodel, and 5) assess the predictive performance. }
        \label{fig:landmaRk}
\end{figure}

We conducted a search of R packages 
available on CRAN that contain the word ``landmark'' in their description. Out of those, six packages were related to time-to-event analysis 
(Supplementary Table \ref{tab:cran-search2-related}). Of these, only \texttt{Landmarking} \citep{barrott_landmarking_2022} corresponds to the setting described in Section \ref{subsec:landmarking}. The latter only considers LOCF and mixed-effects models to summarise longitudinal trajectories, and a Cox PH model for the survival outcomes (it is also possible to use a Fine and Gray model \citep{fine_proportional_1999} in the presence of competing risks). 
These limited options leave a critical gap that precludes the full
potential of landmarking approaches to be realised, leaving users to implement their own
ad-hoc code when applying more flexible models (such as the heterogeneity-aware approach introduced in Section \ref{subsec:heterogeneity}). To address this, we developed \texttt{landmaRk} as a modular and reproducible 
end-to-end framework for dynamic risk prediction using landmarking (Figure \ref{fig:landmaRk}). \texttt{landmaRk} is available on CRAN. 

The \texttt{landmaRk} pipeline 
integrates seamlessly with the \texttt{tidyverse} ecosystem \citep{wickham_welcome_2019}. 
Starting from the definition of risk sets, 
\texttt{landmaRk} supports baseline landmark-specific baseline models without a longitudinal input \citep[which approximates a Cox PH model with time-varying coefficients;][]{van_houwelingen_dynamic_2011}. Our current implementation provides built-in support for LOCF, LME, and LCMM (as per Section \ref{subsec:heterogeneity}) for the longitudinal component, together with a Cox PH model. 
Unlike existing software, the modular specification of \texttt{landmaRk} allows users to extend these options, implementing their own 
definitions for the longitudinal or survival components. 

A key advantage of integrating different model configurations into a common framework is the ability to directly compare predictive performance to inform model choice. 
For example, to assess whether adding longitudinal information improves prediction with respect to baseline landmark-specific survival models. Similarly, users can explore predictive gains linked to the incorporation of latent population heterogeneity (e.g.~LME versus LCMM for the longitudinal model). 
The calculation of predictive performance evaluation metrics is supported in the training set as well as out-of-sample via cross-validation. Cross-validation folds can be set at the start of the pipeline to ensure identical folds are used across different model specifications, 
allowing systematic benchmarking of landmarking approaches. Importantly, the integration of performance evaluation within the \texttt{landmaRk} pipeline ensures that the reported metrics are model agnostic and independent of model-fitting packages. 
That ensures fair performance comparisons, addressing the reproducibility issues highlighted by \cite{sierra_c-index_2025}. 
We consider metrics related to $t$-year dynamic prediction, as per the probabilities in \eqref{eq:dyn_prob}. To assess discrimination, we use the Inverse Probability of Censoring Weighting (IPCW)-adjusted time-dependent area under the ROC curve \citep[tdAUC; ][]{blanche_c-index_2019}.
As a composite measure of discrimination and calibration, we use the IPCW-adjusted Brier score \citep[BS; ][]{graf_assessment_1999}. 
For both metrics, we use the implementation in the \texttt{riskRegression} R package \citep{gerds_riskregression_2025}. Formal definitions for these metrics can be found in Supplementary Section \ref{sec:appendix-metrics}.









\section{Simulation study}
\label{sec:simulation}

\subsection{Data generation}

To evaluate the performance of our proposed heterogeneity-aware landmarking strategy, we consider synthetic data generated under four main scenarios whilst also varying the sample size, yielding eight simulated datasets in total (Table \ref{tab:scenarios}).  
In all cases, the data generating mechanism is consistent with a JLCM specification; that is, the parameters associated to the longitudinal and survival processes are defined by a finite set of $G = 3$ latent classes. 

\begin{table}[h]
\centering
\caption{Simulation settings. All scenarios use $G = 3$ latent classes to define the longitudinal and survival process. The scenarios vary by sample size ($N$), the definition of the survival model (common vs.\ class-specific Weibull shape), and the distribution of the error term associated to the longitudinal process (normal vs.\ heavy-tailed $t_5$).}
\label{tab:scenarios}
\begin{tabular}{cccc}
\hline
\textbf{Scenario} & \textbf{Sample size ($N$)} & \textbf{Weibull Shape} & \textbf{Error distribution} \\
\hline
1 & 1000 & Common         & Normal \\
2 & 2000 & Common         & Normal \\
3 & 1000 & Class-specific & Normal \\
4 & 2000 & Class-specific & Normal \\
5 & 1000 & Common         & $t_5$  \\
6 & 2000 & Common         & $t_5$  \\
7 & 1000 & Class-specific & $t_5$  \\
8 & 2000 & Class-specific & $t_5$  \\
\hline
\end{tabular}
\end{table}

\begin{figure}[h]
\centering
\includegraphics[width=0.85\textwidth]{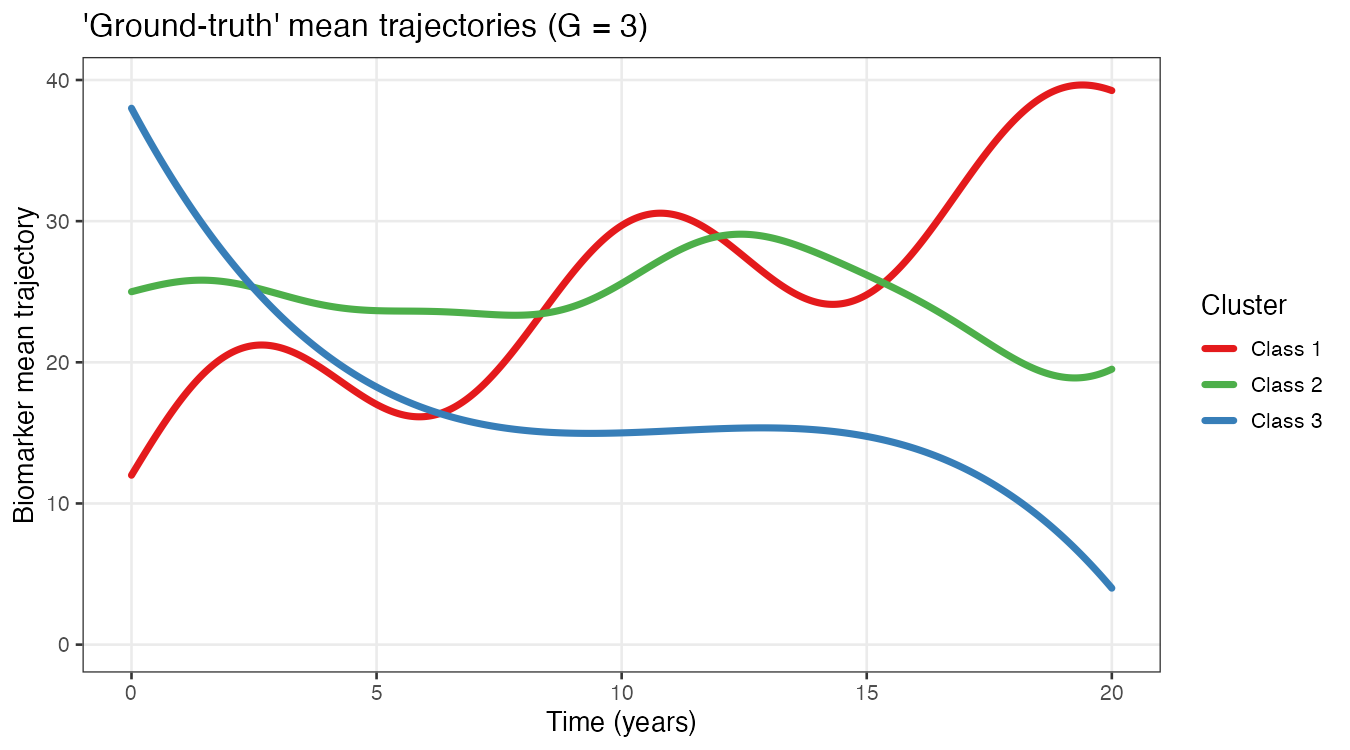}
\caption{Ground-truth mean biomarker trajectories for the $G = 3$ latent clusters used in the simulation study, over a 20-year follow-up. Class~1 (red) shows a U-shape followed by a sharp late increase, Class~2 (green) remains relatively stable with a mid-period peak, and Class~3 (blue) declines steadily from a high baseline.}
\label{fig:mean-trajectories-k3}
\end{figure}

In all scenarios, latent cluster membership is allocated in a probabilistic manner, with
\begin{equation}
    \text{Pr}(c_i = g) = \dfrac{1}{3}, \quad g=1,\dots,3. 
\end{equation}
Cluster-specific mean trajectories are given by \begin{align}
    \textbf{(Cluster 1)} &\quad m_i(t \mid c_i = 1) = 12 + 2t - 0.10t^2 + 0.003t^3 + 5\sin(0.75t) + b_{01i} + 2.5 x_{i} \label{eq:simulated1} \\
    \textbf{(Cluster 2)} &\quad m_i(t \mid c_i = 2) = 25 + 0.5\sin(1.2t) + 0.3t\cos(0.5t) + b_{02i} + 2.5 x_{i} \label{eq:simulated2} \\
    \textbf{(Cluster 3)} &\quad m_i(t \mid c_i = 3) = 38 - 6.5t + 0.60t^2 -0.018t^3 + b_{03i} + 2.5 x_{i} \label{eq:simulated3},
\end{align} where $b_{0gi} \sim \mathcal{N}(0, \sigma^2_{b_0,g})$ are individual-specific random intercepts, with cluster-specific variance parameters given by $(\sigma^2_{b_0,1},\sigma^2_{b_0,2},\sigma^2_{b_0,3}) = (4.0, 3.5, 4.0)$. We include a binary covariate $x$, simulated from a Bernoulli(0.5) distribution. This introduces a fixed-effect which is common across latent classes. 
The time-dependent components linked to each cluster-specific mean trajectory are displayed in Figure \ref{fig:mean-trajectories-k3}. Longitudinal measurements $y_{ij}$ 
conditioned on membership of the $g$th cluster ($c_i = g$) are then defined as \begin{equation}
    Y_{ij} \big|_{c_i = g} = \max\left\{m_i(t \mid c_i = g) + \varepsilon_{ij}, 0.01 \right\} \label{eq:simulated4}, 
\end{equation}
where $\varepsilon_{ij}$ is an error term that represents measurement error, and the left-truncation at $0.01$ emulates a lower limit of detection (often encountered in biomarker measurements). We first consider a Gaussian model for the error term; that is, $\varepsilon_{ij} \sim \mathcal{N}(0, \sigma_{\varepsilon,g}^2)$, with $(\sigma_{\varepsilon,1},\sigma_{\varepsilon,2},\sigma_{\varepsilon,3}) = (1.6, 1.4, 1.6)$. This is the error distribution assumed by the LCMM longitudinal submodel, and therefore represents the scenario in which the measurement error distribution is correctly specified. 
We also consider relaxing this assumption, 
assuming that $\varepsilon_{ij} \sim \sigma_{\varepsilon,g}\ t_5$,
where $t_5$ denotes the standard Student $t$-distribution with 5 degrees of freedom. This scenario imposes a heavy-tail distribution of the error term, 
motivated by the highly noisy nature of certain biomarkers in real-world data.

The survival component is specified with a Weibull distribution, whose parameters also cluster-specific: 
\begin{align} 
    h_i(t \mid c_i = g, w_{i}) = \lambda_g \alpha_g t^{\alpha_g-1} \exp(\gamma_{g} w_{i}) \label{eq:simulated5}
\end{align}
where $w$ is a continuous covariate whose values are simulated from a uniform distribution over $(-1.732, 1.732)$ and $(\gamma_{1}, \gamma_{2}, \gamma_{3}) = (0.08, 0.04, 0.01)$. The Weibull scale parameter, $\lambda_g$, is specified to introduce event rates over the 20-year follow-up period of $\approx$ $0.90$, $0.60$ and $0.30$ for clusters 1, 2 and 3, respectively. Censoring times were sampled from a $\text{Exponential}(0.00256)$ distribution, resulting in approximately a $5\%$ censoring. First, we assume $\alpha_g \equiv \alpha$, where 
$\alpha = 1.5$. In such cases, the shape parameter of the Weibull 
model 
is shared across clusters, and therefore the hazard increases over time at a rate that is common across subjects. 
We also consider scenarios with cluster-specific Weibull shape parameters, 
where $(\alpha_1, \alpha_2, \alpha_3) = 
(1.2, 1.5, 1.9)$. 
In this case, 
the hazard accelerates at different rates across clusters.

\subsection{Methods}

We fitted landmarking models as well as JLCMs to all eight simulated datasets. For the JLCM, the number of clusters is set to $G = 1, 2, 3$ and $4$. For landmarking, we use a static landmark-specific model with no dynamic covariates, as well as different strategies to summarise the longitudinal trajectories: LOCF, LME, and LCMMs with $G = 2, 3$ and $4$ clusters. The longitudinal submodels of both landmarking (LME and LCMM) and JLCMs were specified with fixed and random intercepts, fixed and random slopes for time, a fixed-effect term and an additional fixed effect for the binary covariate $x$. For JLCMs and our proposed heterogeneity-aware landmarking approach, all model parameters were specified to be cluster-specific; that is,
\begin{equation}
    m_i(t \mid c_i = g) = (\beta_{0g} + b_{0gi}) + (\beta_{1g} + b_{1gi}) t + \beta_{2g} x_{i}. \label{eq:model-fit}
\end{equation} Note that the mean trajectory in \eqref{eq:model-fit} (which is linear with respect to time) is purposely misspecified with respect to the data-generating process (which is non-linear). In our main landmarking analyses, a separate longitudinal model was fitted at each landmark time $s$, using only the longitudinal measurements observed prior to that landmark and individuals in the corresponding risk set $\mathcal{R}_s$. As a sensitivity analysis, we also considered an alternative specification in which a single longitudinal model was fitted once using all available longitudinal data, with the resulting 
model then used across all landmark times. In both cases, only longitudinal measurements prior to $s$ where used for prediction.

For landmarking approaches, the survival component is a Cox PH model 
dependent on 
$w$, the predicted value for the longitudinal measurements at time $s$, $\hat{y}_i(s)$, and optionally the predicted cluster label $\hat{c}_i$. That is,
\begin{equation}
    h_i(t | w_i, \hat{y}_i(s), \hat{c}_i) = h_0(t) \exp \left(\gamma w_i + \alpha_1 \hat{y}_i(s) \right)
\end{equation}
or
\begin{equation}
    h_i(t | w_i, \hat{y}_i(s), \hat{c}_i) = h_0(t) \exp \left(\gamma w_i + \alpha_1 \hat{y}_i(s) + \sum\limits_{g=2}^G \alpha_g \mathbf{1}(\hat{c}_i = g)\right).
\end{equation}
The survival component of the JLCMs is a parametric Weibull model with cluster-specific parameters as in \eqref{eq:simulated5}.

In all cases, we assessed predictive performance at $s = 9$ and $s = 12$, whilst considering different prediction horizons (denoted by $w$ in \eqref{eq:dyn_prob}). 
We assessed predictive performance using 5-fold cross-validation, where folds were defined via stratified sampling using the event indicator as the stratifying variable. This is to ensure that the number of events does not vary substantially across cross-validation folds. To ensure a fair comparison, identical folds were used across methods. Performance was assessed on the `pooled' set of out-of-sample predictions, using the tdAUC and 
the BS to asses discrimination and calibration across different prediction horizons $w$ (see Section \ref{sec:landmaRk}), and 95\% confidence intervals were derived using bootstrapping. 

\subsection{Results}

\begin{figure}[H]
  \centering
  \includegraphics[width=\linewidth]{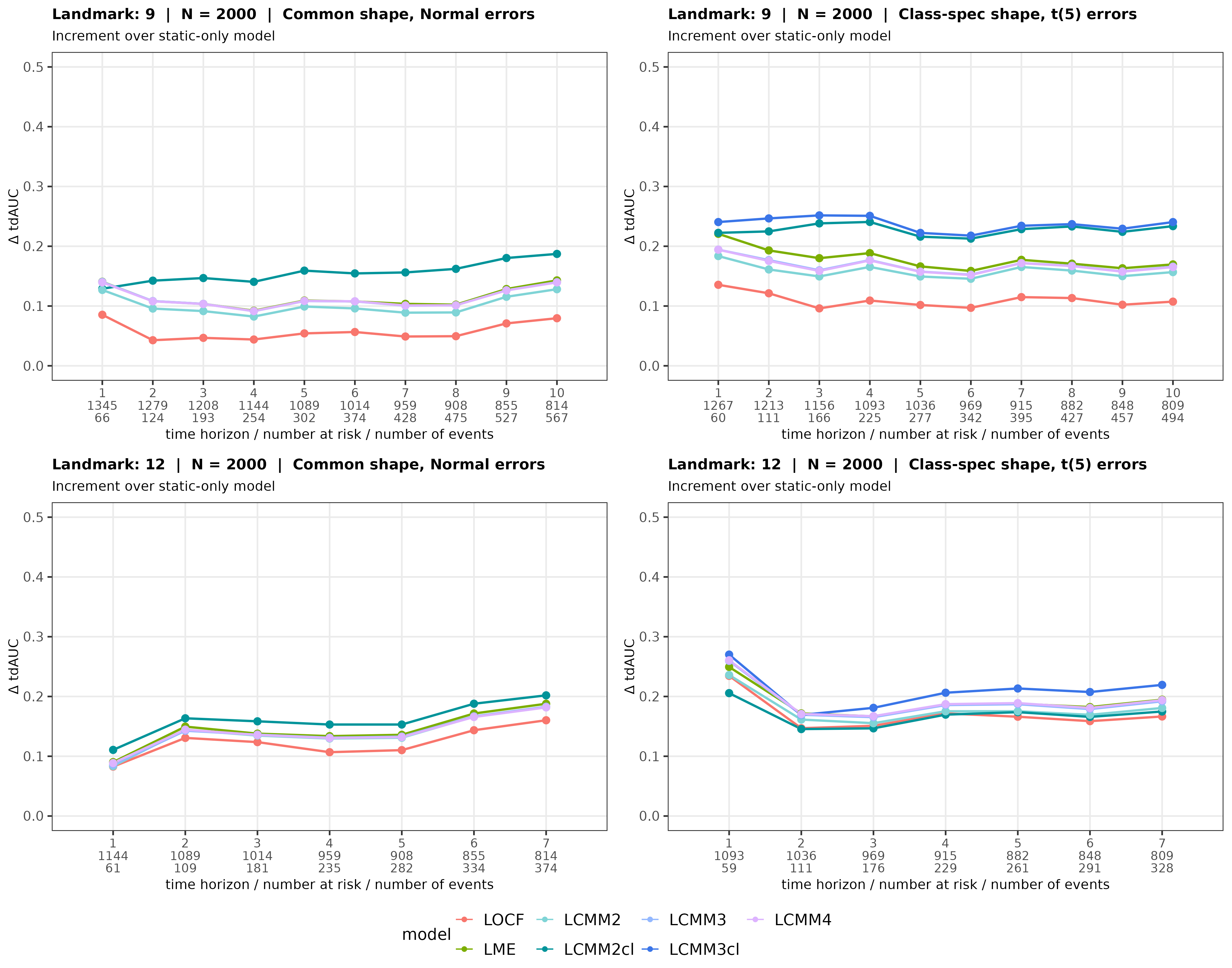}
  \caption{Cross-validated tdAUC increment over the static-only model in selected simulation scenarios at landmarks 9 and 12. Lines show the increment in time-dependent AUC relative to the static-only Cox model, pooled across five cross-validation folds. Only models that converged in all five folds are shown. x-axis labels give the time horizon, number at risk, and number of events. Results are not shown for the models LCMM3cl (N=2000, common Weibull shape, $t_5$ errors) as the optimiser failed to converge in one of the cross-validation training folds.}
  \label{fig:auct-small}
\end{figure}
\begin{figure}[H]
  \centering
  \includegraphics[width=\linewidth]{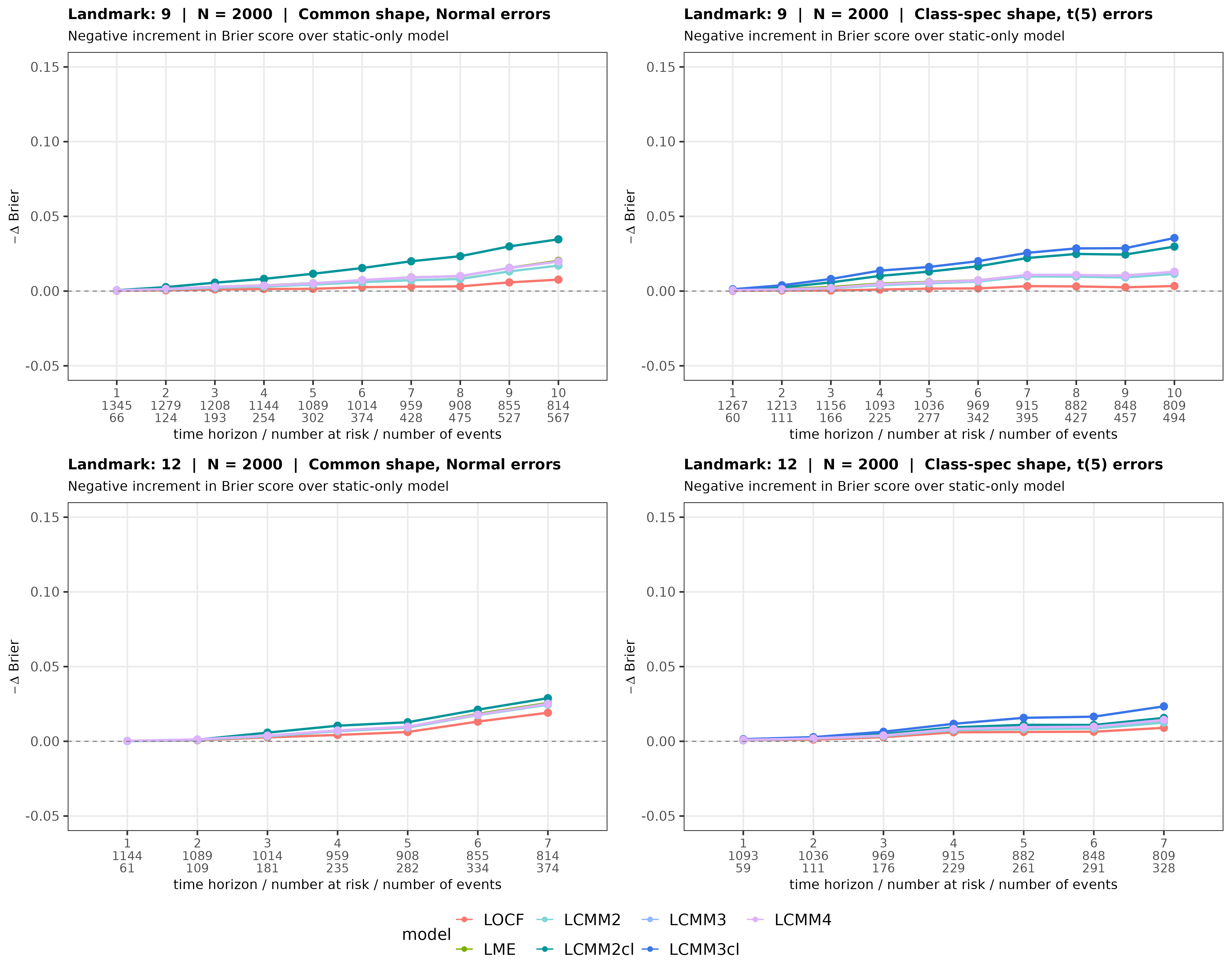}
  \caption{Negative increment in cross-validated Brier score over the static-only model in selected simulation scenarios at landmarks 9 and 12. Positive values indicate improvement over the static-only Cox model. Only models that converged in all five folds are shown. Results are not shown for the models LCMM3cl (N=2000, common Weibull shape, $t_5$ errors) as the optimiser failed to converge in one of the cross-validation training folds.}
  \label{fig:brier-small}
\end{figure}

For all landmarking approaches, Figures \ref{fig:auct-small} and \ref{fig:brier-small} display respectively the estimated increment in tdAUC and BS with respect to a baseline landmark-specific survival model without longitudinal measurements, at selected scenarios and landmark times $s = 9,12$ (Supplementary Figures \ref{fig:auct_lm9}-\ref{fig:brier_lm12} display results for the remaining scenarios, with the corresponding tdAUC and BS estimates shown in  
Supplementary Tables \ref{truncated-tablesauctClassSpect5tex}-\ref{truncated-tablesbrierCommonShapeNormaltex}). 
All landmarking approaches led to an improvement of predictive performance with respect to the baseline models.
The LCMM-based models that exclude the predicted cluster label $\hat{c}_i$ do not improve on LME despite their added complexity, and are sometimes worse. The two are effectively interchangeable in the class-specific-shape scenarios, and in the common-shape scenarios LME is frequently the better of the two. Indeed, $\hat{y}_i(s)$ predictions obtained from LCMM (as per Equation \eqref{eq:summary-lcmm}) are largely similar to the values predicted by LME. In contrast, the heterogeneity-aware landmarking models that include the predicted cluster labels (denoted as e.g.~`LCMM2cl' when number of latent classes is assumed to be $G = 2$; the LCMM4cl specification is excluded from the main analysis as, upon fitting, some clusters were found to be empty at one or more landmark times) are the top performers. 
In most scenarios, discrimination performance of JLCM$G$ is similar to that of the LCMM$G$cl landamrking specifications, for fixed $G$, whereas in terms of calibration the performance of JLCM is substantially worse than that of landmarking. Overall, the performance of LCMM$G$cl is superior to the joint models. Moreover, JLCMs are slower than heterogeneity-aware landmarking specifications (Supplementary Table \ref{tab:running_times_N2000}).

Supplementary Figures \ref{fig:supp-auct_lm9}-\ref{fig:supp-brier_lm12} and Supplementary Tables \ref{nontruncated-tablesauctClassSpect5tex}-\ref{nontruncated-tablesbrierCommonShapeNormaltex} display results for the sensitivity analysis, where a single longitudinal model was fitted using all available longitudinal data. 
Results are largely similar to those discussed above, and the LCMMgcl models remain the strongest. 
However, over-specifying the number of clusters is catastrophic --- most acutely for calibration.
In several scenarios LCMM4cl falls below the static baseline on discrimination while its Brier score worsens dramatically. The collapse is far more severe for calibration than for discrimination. This highlights a key advantage of fitting landmark-specific longitudinal models, which appear to be largely insensitive to over-specification of the number of clusters.


In summary, these results demonstrate that accounting for the presence of latent population heterogeneity can lead to substantial improvements in predictive performance. 
Furthermore, our simulated scenarios suggest that the added value of the our proposed heterogeneity-aware landmarking strategy (over traditional LME-based approaches) mostly lies in incorporating the predicted cluster labels, 
not in the reconstruction of the biomarker trajectory itself. As discussed above, predicted cluster labels provide a interpretable summary for longitudinal trajectories, capturing information about their shape. Beyond considering \emph{where} longitudinal measurements are predicted to be at each landmark time $s$, this captures information about \emph{how} they got there. This is particularly critical in clinical applications, where information about a patient's trajectory can convey important information about future event risk.

\section{Case study: the \texttt{aids} dataset} \label{sec:real-data}

\subsection{Data overview}

We 
next consider the widely known \texttt{aids} dataset, as included in the \texttt{JMbayes2} R package \citep{rizopoulos_jmbayes2_2025}. The dataset relates to 
a clinical trial comparing two antiretroviral drugs, didanosine (ddI) and zalcitabine (ddC). At baseline, CD4 counts were recorded, along with 
other static covariates (Table \ref{tab:baseline}). In four subsequent follow-up visits, CD4 counts were measured, resulting in a total of 1,405 longitudinal measurements across $N = 467$ individuals ($\approx 3$ measurements, on average, per individual). Hereafter, we consider square root-transformed CD4 measurements. Interest lies in leveraging the 
trajectories of 
CD4 counts to predict time to death \citep{abrams_comparative_1994}. Overall, $188$ death events were observed with the remaining $279$ subjects being right censored. 

\begin{figure}[t]
    \centering
        \includegraphics[width=\textwidth]{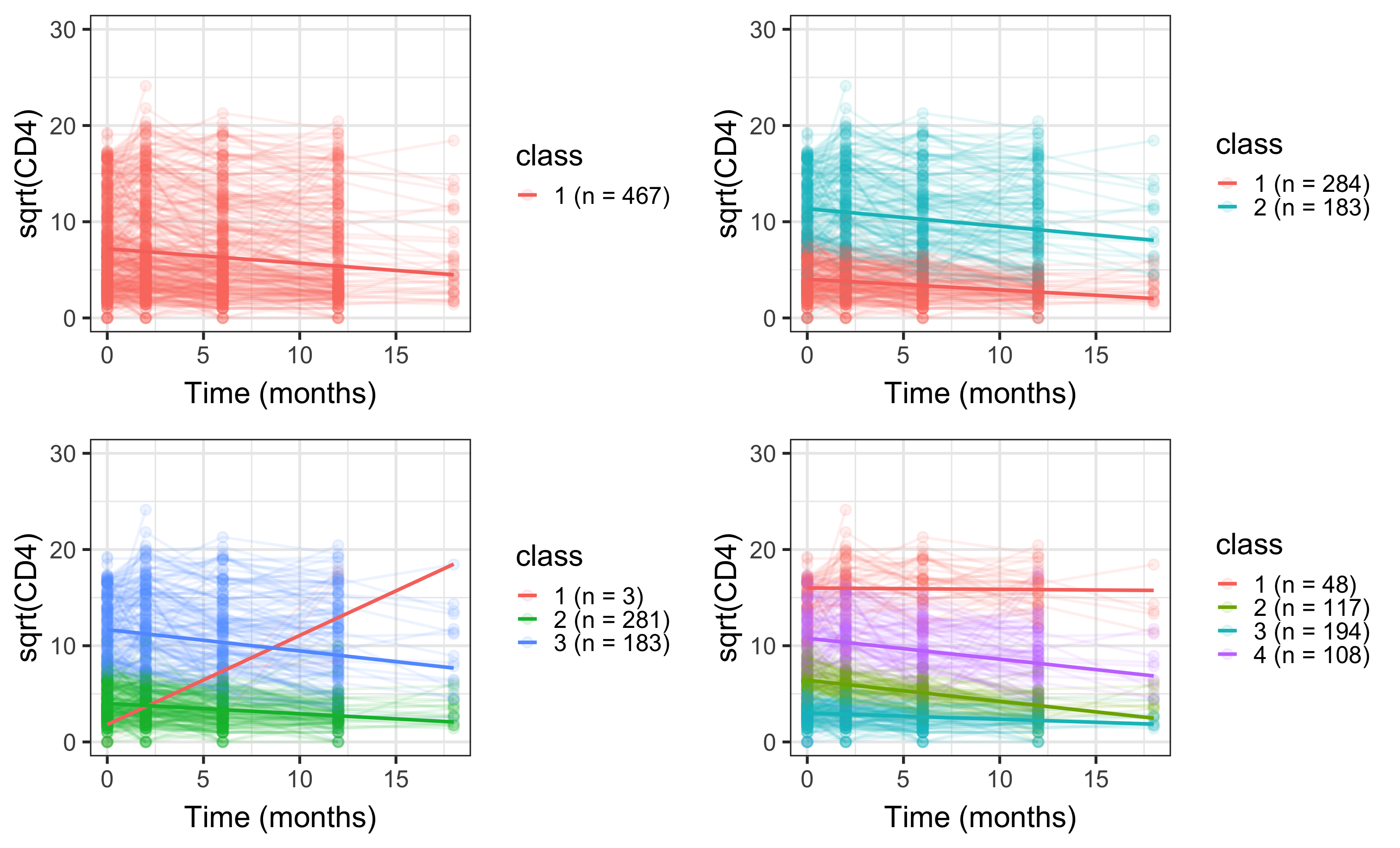}
    \caption{Longitudinal trajectories of square root-transformed CD4 counts in the \texttt{aids} dataset along with cluster membership allocation and estimated mean trajectories, according to LCMM with 1, 2, 3 and 4 clusters. The colours indicate cluster membership. 
    Results indicate the presence of at least two clusters.}
    \label{fig:longitudinal-trajectories}
\end{figure}

The data exhibits substantial heterogeneity in the longitudinal CD4 trajectories. This is partly explained by the available baseline covariates; for example, individuals with a previous AIDS diagnosis (\texttt{prevOI}) have, on average, lower CD4 counts than those without (Supplementary Figure \ref{fig:longitudinal-trajectories-alternative}). Hereafter, baseline covariates are excluded when modelling longitudinal CD4 measurements. That is, their potential effect is treated as latent population heterogeneity. To further explore the structure of the longitudinal trajectories, we fitted LCMM models ($G = 1, \ldots, 6$) to the full dataset (all available measurements, regardless of the event status) as per the model specification in Equation \eqref{eq:longitudinal_aids} (Figure \ref{fig:longitudinal-trajectories}). With $G = 2$, the recovered clusters appear to largely recapitulate the heterogeneity explained by the observed covariate \texttt{prevOI}, whilst the $G = 4$ solution further subdivides these two groups, distinguishing individuals with relatively stable CD4 trajectories from those experiencing a decline. Of note, for $G = 3, 5$ and $6$, the model identified a small cluster of individuals with increasing CD4 trajectories (Supplementary Figure \ref{fig:lcmm5and6}). 


\subsection{Methods}


For landmarking, we used the following specifications for the longitudinal component: LOCF, LME, and our proposed heterogeneity-aware landmarking approach using LCMMs with 2, 3 and 4 clusters. As a baseline, we also consider a landmark-specific model that does not consider longitudinal 
CD4 measurements. 
The longitudinal mean trajectory of square root-transformed CD4 counts is 
assumed to be
\begin{equation} \label{eq:longitudinal_aids}
    m_i(t \mid c_i = g) = \beta_{0g} + b_{0gi} + (\beta_{1g} + b_{1gi}) t,
\end{equation} where $\beta_{0g}$ and $\beta_{1g}$
are cluster-specific fixed effects, and $(b_{0gi}, b_{1gi})^\top \sim \mathcal{N}(\mathbf{0}, \Sigma_g)$ denote cluster-specific random effects, with $\Sigma_g$ being the random-effect variance-covariance matrix. The specification of the temporal trend follows \cite[\S5.4]{rizopoulos_joint_2012}, with \texttt{drug} excluded from the longitudinal submodel to treat its potential effect as latent population heterogeneity. Longitudinal measurements are modelled with 
\begin{equation} \label{eq:longitudinal-aids-error}
    y_{ij} \big|_{c_i=g} = y_i(t_{ij} \mid c_i = g) = m_i(t \mid c_i = g) + \varepsilon_{ij},
\end{equation}
where $\varepsilon_{ij} \sim \mathcal{N}(0, \sigma^2)$.
The survival component of the landmarking approaches is a Cox PH model characterised a hazard function dependent on the static covariate $\texttt{drug}_i$, the predicted value for the dynamic covariate at time $s$, $\hat{y}_i(s)$, and optionally the predicted cluster label $\hat{c}_i$. That is,
\begin{equation}
    h_i(t | \texttt{drug}_i, \hat{y}_i(s)) = h_0(t) \exp \left(\gamma_1 \texttt{drug}_i + \alpha \hat{y}_i(s)\right), \label{eq:survival-lcmm1}
\end{equation} 
or 
\begin{equation}
    h_i(t | \texttt{drug}_i, \hat{y}_i(s), \hat{c}_i) = h_0(t) \exp \left(\gamma_1 \texttt{drug}_i + \alpha_1 \hat{y}_i(s) + \sum\limits_{g=2}^G \alpha_g \mathbf{1}(\hat{c}_i = g)\right),\label{eq:survival-lcmm2}
\end{equation} 

Hereafter, landmarking models with LCMM as longitudinal component and \eqref{eq:survival-lcmm1} as the survival component are referred to as LCMM$G$, whereas those with \eqref{eq:survival-lcmm2} as survival component are referred to as LCMM$G$cl. Due to the limited sample size ($N = 467$), we only consider the inclusion of the predicted cluster labels for $G=2$. As an alternative approach that also incorporates latent heterogeneity, we applied JLCM. 
For this purpose, we used the implementation in the \texttt{lcmm} R package \citep{proust-lima_estimation_2017}, and the number of clusters was set to $G = 1, 2, 3$ and $4$. The longitudinal component was defined as in \eqref{eq:longitudinal_aids}. The survival component of the JLCMs is a parametric Weibull model with hazard function with cluster-specific baseline hazards $h_{0g}(t)$:
\begin{equation}
    h_i(t | c_i = g,  \texttt{drug}_i, \hat{y}_i(s)) = h_{0g}(t) \exp \left(\gamma_{1g} \texttt{drug}_i\right).
\end{equation}

In all cases, we considered predictions based on longitudinal measurements available up to landmark times $s=7,8,\dots,12$ months after baseline, whilst varying the prediction horizon $w$.
For all models, we assessed predictive performance using 5-fold cross-validation following the same procedure as in Section \ref{sec:simulation}.

\begin{table}[t]
\setlength{\textfloatsep}{12pt plus 2pt minus 4pt}
\caption{Baseline characteristics of patients in the dataset \texttt{aids}. \texttt{prevOI} indicates previous diagnosis with Acquired Immune Deficiency Syndrome (coded `AIDS'); \texttt{AZT} indicates zidovudine (AZT) intolerance and \texttt{CD4} indicates square root-transformed CD4 at baseline. The two treatment arms are didanosine (ddI) and zalcitabine (ddC).}
    \centering
    \label{tab:baseline}
\begin{tabular}{|l|c|c|c|}
\hline
 & ddC (N=237) & ddI (N=230) & Total (N=467)\\
\hline
\textbf{gender} &  &  & \\
\hline
~~~female & 23 (9.7\%) & 22 (9.6\%) & 45 (9.6\%)\\
\hline
~~~male & 214 (90.3\%) & 208 (90.4\%) & 422 (90.4\%)\\
\hline
\textbf{prevOI} &  &  & \\
\hline
~~~noAIDS & 79 (33.3\%) & 81 (35.2\%) & 160 (34.3\%)\\
\hline
~~~AIDS & 158 (66.7\%) & 149 (64.8\%) & 307 (65.7\%)\\
\hline
\textbf{AZT} &  &  & \\
\hline
~~~intolerance & 146 (61.6\%) & 146 (63.5\%) & 292 (62.5\%)\\
\hline
~~~failure & 91 (38.4\%) & 84 (36.5\%) & 175 (37.5\%)\\
\hline
\textbf{CD4} &  &  & \\
\hline
~~~Mean (SD) & 7.024 (4.652) & 7.238 (4.777) & 7.129 (4.710)\\
\hline
~~~Range & 0.000 - 19.053 & 0.000 - 19.235 & 0.000 - 19.235\\
\hline
\end{tabular}

\end{table}

\subsection{Results}



Tables \ref{tab:auct-aids} and \ref{tab:brier-aids} report cross-validated performance metrics (tdAUC, BS) for the different landmarking and JLCM specifications under consideration, across a set of (landmark $s$, horizon $w$) pairs. Discrimination performance of the joint models is inferior to that of landmarking approaches in most cases, whereas in terms of calibration the performance of JLCM is substantially worse than that of landmarking. Overall, discrimination is highest between 12 to 15 months post-baseline (Supplementary Figure \ref{fig:auct-performance-absolute}). Whilst tdAUC reveals substantial differences in performance across methods, all landmarking specifications lead to largely similar Brier scores (Supplementary Figure \ref{fig:brier-performance-absolute}). 

\begin{figure}[h]
    \centering
    \includegraphics[width=\textwidth]{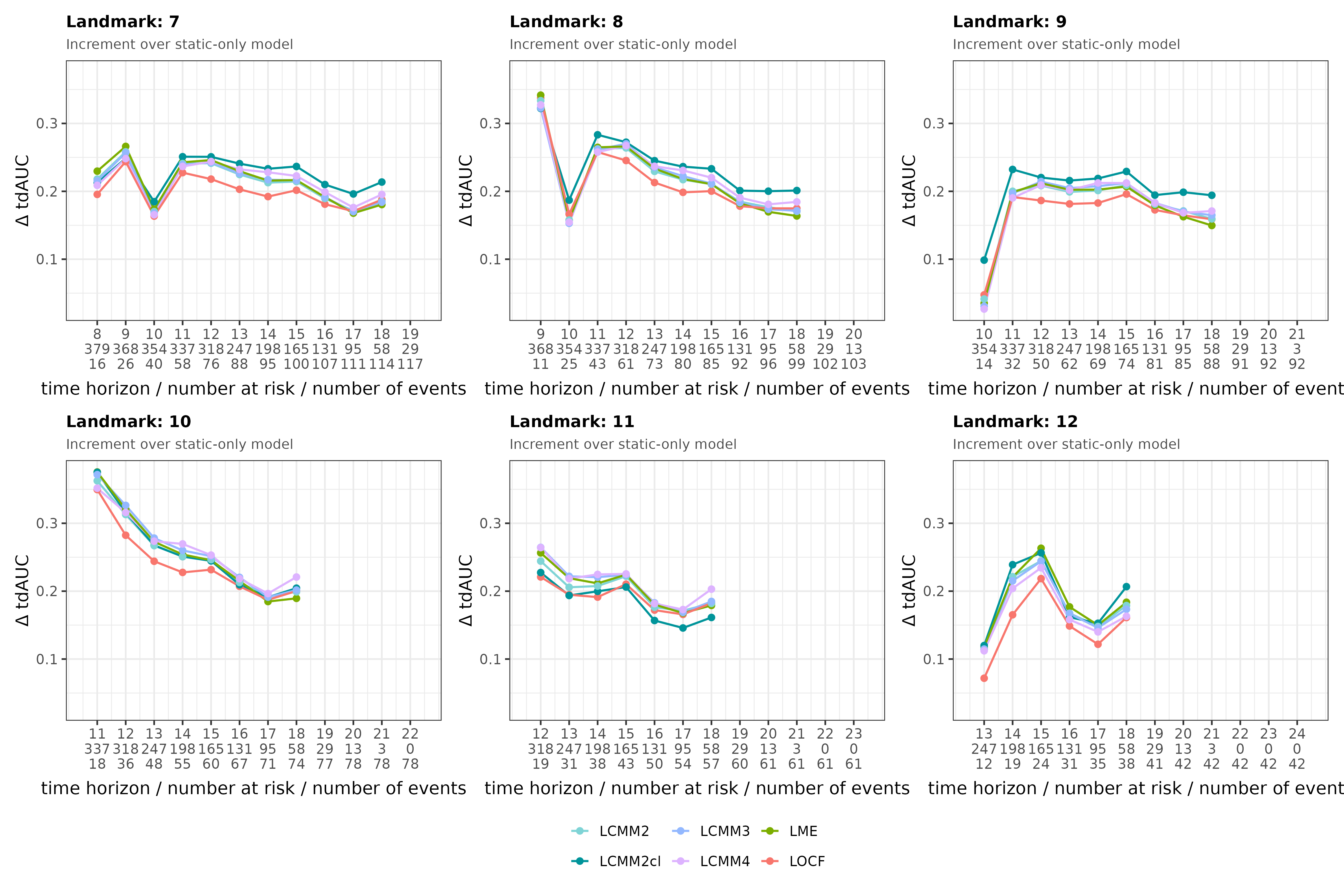}
    \caption{Cross-validated increment in tdAUC over the static-only model for the different model specifications in the landmarking framework for the \texttt{aids} case study. For each specification, the increment in tdAUC is computed on the pooled out-of-sample predictions across the five cross-validation folds.}
    \label{fig:auct-performance}
\end{figure}

\begin{figure}[h]
    \centering
    \includegraphics[width=\textwidth]{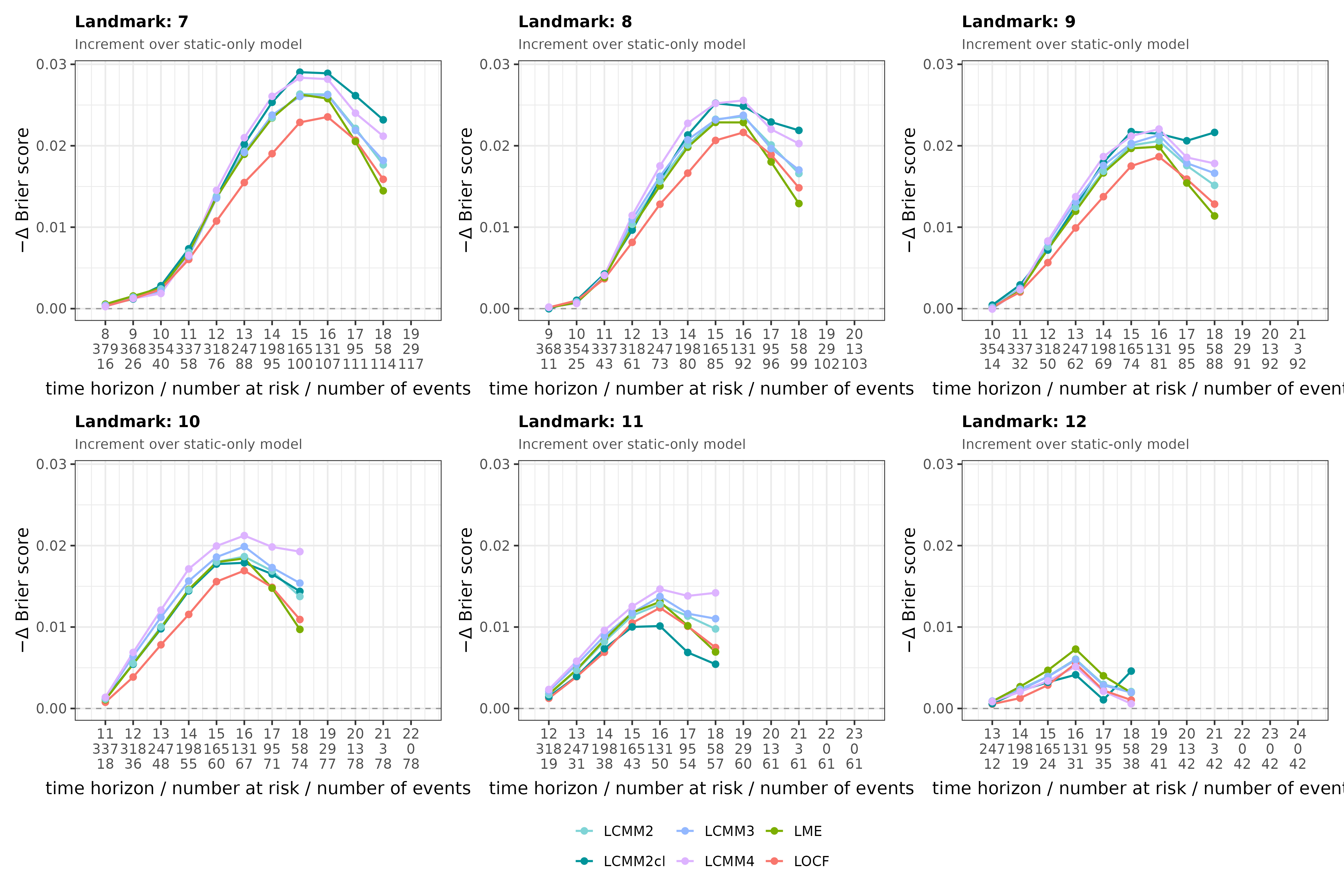}
        \caption{Cross-validated increment in BS over the static-only model for the different model specifications in the landmarking framework for the \texttt{aids} case study. For each specification, the increment in BS is computed on the pooled out-of-sample predictions across the five cross-validation folds.}
    \label{fig:brier-performance}
\end{figure}

Among landmarking specifications, incorporating longitudinal CD4 measurements consistently improves both discrimination and calibration over the static-only model, but the gains are generally lower for LOCF. Using the static-only model as a reference, Figures \ref{fig:auct-performance} and \ref{fig:brier-performance} report the increment in cross-validated performance 
for LOCF, LME and LCMM-based models at different landmarks. 
The best performing model varies across landmarks  and prediction horizons, and the chosen metric (tdAUC or BS) ---  no single model is uniformly best. 

Our heterogeneity-aware landmarking strategy outperforms LOCF and LME specifications in most cases, but this performance strongly depends on the choice of $G$ and on whether the predicted cluster labels are used as an input for the survival model. At the earlier landmarks $(s = 7,8,9)$, 
LCMM2cl leads to the highest tdAUC but its performance deteriorates later on. In contrast, the models that assume three or four clusters (LCMM3 and LCMM4, respectively) have a  more consistent strong performance.
At the final landmark ($s = 12$), LME generally outperforms all the LCMM specifications, 
possibly because 
the risk set becomes smaller, 
resulting in less stable LCMM estimates. 




\begin{landscape}
\begin{table}
\tiny
\caption{Cross-validated tdAUC (95\% bootstrap CI) for all model specifications for the \texttt{aids} case study.}
\centering

\begin{tabular}[t]{ccrrrrrrrrrrrrr}
\toprule
$s$ & $w$ & Static-only & LOCF & LME & LCMM2 & LCMM3 & LCMM4 & LCMM2cl & JLCM1 & JLCM2 & JLCM3 & JLCM4
\\
\midrule
8 & 3 & 0.395 & 0.653 & 0.660 & 0.657 & 0.656 & 0.653 & \textbf{0.678} & 0.497 & 0.633 & 0.644 & 0.640 \\
 &  & (0.307, 0.475) & (0.583, 0.704) & (0.601, 0.721) & (0.589, 0.726) & (0.588, 0.723) & (0.573, 0.713) & \textbf{(0.632, 0.749)} & (0.429, 0.569) & (0.550, 0.694) & (0.571, 0.708) & (0.572, 0.720) \\
8 & 6 & 0.521 & 0.719 & 0.739 & 0.738 & \textbf{0.743} & \textbf{0.751} & \textbf{0.757} & 0.551 & 0.684 & 0.721 & 0.698 \\
 &  & (0.458, 0.583) & (0.645, 0.769) & (0.685, 0.794) & (0.683, 0.800) & \textbf{(0.682, 0.786)} & \textbf{(0.696, 0.809)} & \textbf{(0.715, 0.814)} & (0.486, 0.640) & (0.616, 0.750) & (0.668, 0.778) & (0.642, 0.755) \\
\midrule
10 & 3 & 0.486 & 0.730 & 0.759 & 0.753 & \textbf{0.765} & \textbf{0.760} & 0.754 & 0.555 & 0.722 & 0.753 & 0.715 \\
 &  & (0.395, 0.584) & (0.665, 0.802) & (0.701, 0.814) & (0.699, 0.802) & \textbf{(0.711, 0.833)} & \textbf{(0.701, 0.816)} & (0.686, 0.824) & (0.468, 0.624) & (0.667, 0.790) & (0.692, 0.800) & (0.634, 0.764) \\
10 & 6 & 0.525 & 0.732 & 0.740 & 0.737 & \textbf{0.745} & \textbf{0.743} & 0.734 & 0.578 & 0.709 & 0.722 & 0.657 \\
 &  & (0.451, 0.601) & (0.666, 0.792) & (0.682, 0.809) & (0.676, 0.810) & \textbf{(0.680, 0.811)} & \textbf{(0.667, 0.800)} & (0.672, 0.801) & (0.506, 0.640) & (0.642, 0.779) & (0.668, 0.788) & (0.583, 0.720) \\
\midrule
12 & 3 & 0.461 & 0.679 & 0.724 & 0.705 & 0.705 & 0.695 & 0.717 & 0.618 & 0.701 & \textbf{0.736} & 0.690 \\
 &  & (0.338, 0.587) & (0.549, 0.786) & (0.610, 0.836) & (0.619, 0.803) & (0.607, 0.806) & (0.619, 0.797) & (0.648, 0.792) & (0.502, 0.707) & (0.565, 0.842) & \textbf{(0.652, 0.848)} & (0.572, 0.788) \\
12 & 6 & 0.404 & 0.565 & 0.588 & 0.583 & 0.577 & 0.567 & \textbf{0.611} & 0.511 & 0.580 & \textbf{0.605} & 0.585 \\
 &  & (0.295, 0.520) & (0.444, 0.666) & (0.467, 0.705) & (0.490, 0.695) & (0.469, 0.681) & (0.485, 0.677) & \textbf{(0.497, 0.706)} & (0.393, 0.618) & (0.494, 0.689) & \textbf{(0.501, 0.719)} & (0.482, 0.707) \\\bottomrule
\end{tabular}

\label{tab:auct-aids}

\end{table}
\begin{table}
\tiny
\caption{Cross-validated Brier score (95\% bootstrap CI) for all model specifications for the \texttt{aids} case study.}
\centering
\begin{tabular}[t]{ccrrrrrrrrrrrrr}
\toprule
$s$ & $w$ & Static-only & LOCF & LME & LCMM2 & LCMM3 & LCMM4 & LCMM2cl & JLCM1 & JLCM2 & JLCM3 & JLCM4
\\
\midrule
8 & 3 & 0.102 & 0.099 & 0.098 & \textbf{0.098} & \textbf{0.098} & \textbf{0.098} & \textbf{0.098} & 0.169 & 0.176 & 0.178 & 0.186 \\
 &  & (0.078, 0.122) & (0.080, 0.119) & (0.077, 0.127) & \textbf{(0.075, 0.121)} & \textbf{(0.076, 0.126)} & \textbf{(0.071, 0.120)} & \textbf{(0.078, 0.129)} & (0.161, 0.177) & (0.164, 0.186) & (0.165, 0.193) & (0.173, 0.197) \\
8 & 6 & 0.172 & 0.155 & 0.152 & \textbf{0.152} & \textbf{0.151} & \textbf{0.149} & \textbf{0.150} & 0.229 & 0.216 & 0.221 & 0.232 \\
 &  & (0.145, 0.191) & (0.137, 0.178) & (0.130, 0.176) & \textbf{(0.136, 0.168)} & \textbf{(0.135, 0.171)} & \textbf{(0.128, 0.167)} & \textbf{(0.133, 0.172)} & (0.224, 0.234) & (0.198, 0.233) & (0.204, 0.240) & (0.211, 0.250) \\
\midrule
10 & 3 & 0.120 & 0.112 & 0.110 & \textbf{0.110} & \textbf{0.109} & \textbf{0.108} & 0.110 & 0.215 & 0.206 & 0.221 & 0.232 \\
 &  & (0.091, 0.141) & (0.087, 0.134) & (0.084, 0.133) & \textbf{(0.089, 0.128)} & \textbf{(0.085, 0.133)} & \textbf{(0.084, 0.126)} & (0.080, 0.130) & (0.210, 0.219) & (0.191, 0.225) & (0.207, 0.239) & (0.210, 0.252) \\
10 & 6 & 0.174 & 0.157 & 0.156 & \textbf{0.155} & \textbf{0.154} & \textbf{0.153} & 0.156 & 0.264 & 0.247 & 0.277 & 0.289 \\
 &  & (0.147, 0.200) & (0.134, 0.182) & (0.130, 0.179) & \textbf{(0.132, 0.181)} & \textbf{(0.132, 0.174)} & \textbf{(0.132, 0.182)} & (0.128, 0.175) & (0.258, 0.271) & (0.218, 0.277) & (0.250, 0.305) & (0.254, 0.322) \\
\midrule
12 & 3 & 0.088 & 0.085 & 0.083 & 0.084 & 0.084 & 0.084 & 0.084 & 0.264 & 0.267 & 0.309 & 0.316 \\
 &  & (0.056, 0.118) & (0.057, 0.111) & (0.054, 0.116) & (0.060, 0.111) & (0.062, 0.112) & (0.058, 0.111) & (0.059, 0.114) & (0.258, 0.270) & (0.235, 0.299) & (0.279, 0.337) & (0.284, 0.346) \\
12 & 6 & 0.162 & 0.161 & 0.160 & \textbf{0.160} & \textbf{0.160} & 0.161 & \textbf{0.157} & 0.318 & 0.323 & 0.374 & 0.363 \\
 &  & (0.121, 0.195) & (0.128, 0.198) & (0.123, 0.203) & \textbf{(0.126, 0.193)} & \textbf{(0.131, 0.199)} & (0.119, 0.196) & \textbf{(0.123, 0.193)} & (0.301, 0.331) & (0.279, 0.374) & (0.316, 0.439) & (0.310, 0.426) \\
 \bottomrule
\end{tabular}

\label{tab:brier-aids}
\end{table}
\end{landscape}

\section{Conclusions}
\label{sec:conclusions}

Our contributions are twofold. First, we introduce a novel, computationally efficient, and heterogeneity-aware method for dynamic prediction of time-to-event outcomes. Our approach builds upon the flexibility of landmarking, and the ability to capture latent population structure that is provided by LCMM. Secondly, we develop \texttt{landmaRk} as modular software that provides an end-to-end framework for landmarking analyses, whilst allowing flexible configurations for the longitudinal and survival components. By standardising
model outputs into a common format, this aids comparison between different model choices, accelerating model development and performance evaluation. In combination, these contributions will support reproducibility in the development of dynamic risk prediction strategies, whilst accounting for the population heterogeneity that is often encountered in EHR settings.  

Existing landmarking approaches either ignore measurement error in longitudinal measurements 
or assume a homogeneous distribution of the latent trajectories underpinning those measurements, whereas joint modelling approaches that account for latent heterogeneity can be computationally prohibitive in EHR settings. We overcame these limitations by extending the landmarking paradigm with an LCMM for the longitudinal 
component. Our simulations and case study underscore  the importance of explicitly accounting for potential latent population heterogeneity, demonstrating that our proposed heterogeneity-aware method can improve prediction performance (compared to existing landmarking alternatives) in the presence of such sub-population structure. 
Importantly, the modularity of our approach trivially enables parallelisation, making it suitable to the sample sizes 
found in EHR settings.


An important limitation to our method is that the number of clusters $G$ has to be specified in advance, or chosen via a grid-search. Selecting the optimal number of clusters is not straightforward given the nature of landmarking, where multiple landmark-specific models are to be fitted sequentially. 
In practice, an exploratory analysis of the full dataset (all longitudinal measurements for all individuals, regardless of their event status) may be used to select candidate values for $G$. A final choice for $G$ and other model specification aspects (e.g.~whether to include predicted cluster labels in the survival model) may be made post-hoc based on predictive performance. For that purpose, it is critical to consider metrics that are consistent with the desired use-case \citep{lillelund_position_2026}. Other practical considerations (e.g.~sample size, running times) may also be taken into account. 
Importantly, our results suggest that these challenges are attenuated by robustness to slight misspecification of the number of latent classes. Indeed, if the primary aim is to perform prediction, $G$ may be considered as a nuisance parameter. 

A general limitation of the two-step landmarking framework is that it ignores uncertainty in the prediction of longitudinal outcomes. In our context, uncertainty also arises on the predicted cluster labels. 
This may be particularly important at earlier landmarks or for individuals for which the number of longitudinal measurements is small. Uncertainty in cluster allocations is taken into account when averaging cluster-specific predictions. However, such uncertainty is ignored when including predicted cluster labels in the survival model. A possible alternative would be to consider the predicted probabilities, as opposed to the cluster labels themselves.

Finally, there are limitations associated to the use of LCMM. First, it assumes that the error term in the longitudinal component is Gaussian, 
and therefore will not adequately handle the heavy-tailed residual variation often encountered in real-world biomarker data. 
Moreover, whilst our approach is more scalable than JLCM, 
some computational challenges remain. This includes possible convergence issues at earlier landmarks for which less longitudinal follow-up is available or where the risk sets become too small. Scalability is also challenging in high-frequency settings, where the number of longitudinal measurements is large. 
Alternative approaches for modelling heterogeneous longitudinal trajectories such as generalised additive models \citep{zhang_time-varying_2013} and latent Gaussian processes \citep{hall_modelling_2008}, may offer a viable option. 
Importantly, the modular design of the \texttt{landmaRk} package enables easy replacement of the longitudinal component, which will facilitate further development of the landmarking paradigm 
as new methodological needs arise from applied research.

In sum, we have introduced a heterogeneity-aware landmarking approach and a modular \texttt{R} package that together provide a computationally efficient framework for dynamic risk prediction in the presence of latent population heterogeneity, as commonly encountered in EHR settings. It is our hope that the modular design of \texttt{landmaRk} will encourage further methodological developments of the landmarking paradigm as informed by practical applications, whilst the built-in performance evaluation tools will facilitate systematic benchmarking across model specifications.
\section{Software and reproducibility}
\label{sec:reproducibility}

All analyses were performed in R v4.4.1. Packages used in the analysis were \texttt{lme4} v1.1.37, \texttt{lcmm} v2.2.1 and \texttt{survival} v3.6.4. Packages used for performance evaluation were \texttt{timeROC} v0.4 and \texttt{pec} v2025.6.24. The analysis in Section \ref{sec:real-data} is fully reproducible and publicly available at \url{https://github.com/VallejosGroup/landmarking_case_study_aids}. The analysis in Section \ref{sec:simulation} is fully reproducible and publicly available at \url{https://github.com/VallejosGroup/landmarking_simulated_data}.

\clearpage

\section*{Acknowledgments}

VV-P, NC-C, and CWL are funded by UK Research and Innovation via a Future Leaders Fellowship ‘Predicting outcomes in IBD’ (MR/S034919/1) awarded to CWL as Chief Investigator.

The authors thank Begoña Bolós Sierra for her contributions to the \texttt{landmaRk} package and discussions on performance metrics, and Evangelin Shaloom Vitus for her extensive troubleshooting support during the development of the package.

\bibliographystyle{biorefs}
\bibliography{refs}

\newpage
\appendix
\renewcommand{\thefigure}{S\arabic{figure}}
\renewcommand{\thetable}{S\arabic{table}}
\setcounter{figure}{0}
\setcounter{table}{0}
\setcounter{section}{18}
\section{Supplementary Material}

\label{sec::suplementary}

\subsection{Supplementary Methods}
\subsubsection{LCMM model specification details} \label{sec:lcmm-logistic-regression}

The vector of covariates is partitioned as $\textbf{x}_i = (\textbf{x}_{1i}, \textbf{x}_{2i})$, where $\textbf{x}_{1i}$ informs cluster allocations and $\textbf{x}_{2i}$ informs longitudinal trajectories. The number of clusters, $G$, is specified in advance, and 

\begin{equation}
    \pi_{ig} = \text{Pr}(c_i = g \mid \textbf{x}_{1i}) = \dfrac{\exp \left( \xi_{0g} + \textbf{x}_{1i}^T \boldsymbol{\xi}_{1g} \right)}{\sum_{\ell = 1}^G \exp \left( \xi_{0\ell} + \textbf{x}_{1i}^T \boldsymbol{\xi}_{1\ell} \right)}, \quad i = 1, \ldots, N; g = 1, \ldots, G,
\end{equation} where 
$\{\xi_{og}, \boldsymbol{\xi}_{1g}, g = 1, \ldots G \}$ are unknown parameters. For identifiability, the constraint $\xi_{0G} = 0,  \boldsymbol{\xi}_{1G} = \boldsymbol{0}$ is imposed.

Conditioned on membership of cluster $g$, observations $y_{ij} = y_i(t_{ij})$ are assumed to be noisy measurements of the covariate process associated with its cluster, $c_i$, as specified in (\ref{eq:cluster_lme}). The latter, $m_i(t \mid c_i=g)$, is further into decomposed into fixed and random components as \begin{equation} \label{eq:fixed-process}
    m_i(t \mid \mathbf{x}_{i}, c_i=g) = m^F_i(t \mid \mathbf{x}_{i}, c_i = g) + m^R_i(t \mid \mathbf{x}_{i}, c_i = g),
\end{equation}
where \begin{equation}
    m^F_i(t \mid \mathbf{x}_{i}, c_i = g) = \boldsymbol{\beta}_g^\top \textbf{x}_{2i}(t) 
\end{equation} captures the mean cluster-specific trajectory and \begin{equation} \label{eq:random-process}
    m^R_i(t \mid \mathbf{x}_{i}, c_i = g) = \textbf{b}_{ig}^\top \textbf{z}_i(t), \quad \mathbf{b}_{ig} \sim \mathcal{N}(\mathbf{0}, \mathbf{\Sigma}_g),
\end{equation} captures inter-individual variation within each cluster. In equations \ref{eq:fixed-process} and \ref{eq:random-process}, $\mathbf{x}_{2i}(t)$ denotes a vector of covariates typically including an intercept term, $\mathbf{x}_{2i}$ and $t$ (or additional time-dependent fixed effects, e.g. defined using natural cubic splines), $\mathbf{z}_i(t)$ is a subset of $\mathbf{x}_{2i}(t)$, $\boldsymbol{\beta}_g$ denotes the cluster-specific fixed effects and $\mathbf{\Sigma}_g$ denotes the variance-covariance matrix of the cluster-specific random effects.

\subsubsection{Performance metrics}
\label{sec:appendix-metrics}
The time-dependent area under the ROC curve (tdROC) adjusted for Inverse Probability of Censoring Weights (IPCW) \citep{blanche_estimating_2013}, is defined by 
\begin{equation}
    \widehat{\text{tdAUC}}(s+w \mid s) = \dfrac{\sum\limits_{i=1}^{N_s} \sum\limits_{j=1}^{N_s} \mathbf{1}(s< T_i \le s+w < T_j) \mathbf{1}(\text{Risk}_i>\text{Risk}_j)\dfrac{D_i}{N_s^2 \hat{C}(T_i \mid s) \hat{C}(s+w\mid s)}}{\hat{W}(s+w\mid s)(1-\hat{W}(s+w\mid s))},
\end{equation}
where $\text{Risk}_i$ denotes a risk score or prognostic marker (linear predictor in the Cox PH model), $N_s$ is the size of $\mathcal{R}_s$, $\hat{C}(s+w \mid s)$ is the Kaplan-Meier estimator of the survival function for the censoring process and $1-\hat{W}(s+w\mid s) = \sum_{i=1}^{N_s} \mathbf{1}(s < T \le s+w) \tfrac{D_i}{N\hat{C}(s+w\mid s)}$. 

The IPCW-adjusted Brier score \citep{graf_assessment_1999} is defined by 
\begin{equation}
    \widehat{\text{Brier}}(s+w\mid s) = \dfrac{1}{N_s} \sum\limits_{i=1}^{N_s} \hat{\pi}_i(s+w \mid s)^2 \dfrac{\mathbf{1}(s < T_i \le s+w, D_i = 1)}{\hat{W}(T_i)} + (1-\hat{\pi}_i(s+w \mid s))^2 \dfrac{\mathbf{1}(T_i > s+w)}{\hat{W}(T_i)}
\end{equation}
\subsubsection{CRAN package search}


\begin{sidewaystable}[htbp]
\centering
\small
\setlength{\tabcolsep}{4pt}
\renewcommand{\arraystretch}{1.1}

\caption{%
R packages in CRAN containing the word ``landmark'' in the description and related to landmarking analysis of time-to-event data.
Abbreviations: PH = proportional hazards; KM = Kaplan--Meier; GLM = generalized linear model;
LOCF = last observation carried forward; LME = linear mixed-effects model;
IPTW = inverse probability of treatment weighting;
IPCW = inverse probability of censoring weighting. Five additional packages, unrelated to time-to-event analysis and containing the word ``landmark'' in their description were found (not listed here). Data is up-to-date on 30 January 2026.
}

\begin{tabularx}{\textwidth}
{|>{\raggedright\arraybackslash}p{3.2cm}|X|c|c|X|}
\hline
Name & Description & Longitudinal & Survival & Notes \\ \hline

landest~\citep{parast_landest_2023} &
Landmark Estimation of Survival and Treatment Effect &
None &
KM &
IPTW adjustment \\ \hline

Landmarking~\citep{barrott_landmarking_2022} &
Analysis using Landmark Models &
LOCF, LME &
Cox PH, Fine--Gray &
IPCW-adjusted Brier score and C-index \\ \hline

landmix~\citep{garcia_landmix_2022} &
Prediction for Mixture Data &
None &
Kernel Nelson--Aalen &
 \\ \hline

landmulti~\citep{li_landmulti_2023} &
Prediction with Multiple Short-Term Events &
None &
Varying coefficient GLM &
 \\ \hline

landpred~\citep{huynh_landpred_2025} &
Prediction of a Survival Outcome &
None &
KM &
 \\ \hline

presmoothedTP~\citep{soutinho_presmoothedtp_2025} &
Landmark Aalen--Johansen estimator of transition probabilities for complex multi-state models &
None &
Aalen--Johansen &
 \\ \hline

\end{tabularx}
\label{tab:cran-search2-related}
\end{sidewaystable}

\clearpage


\subsubsection{Results: simulated data (main analysis)}
\newpage
\begin{figure}[H]
  \centering
  \includegraphics[width=\linewidth]{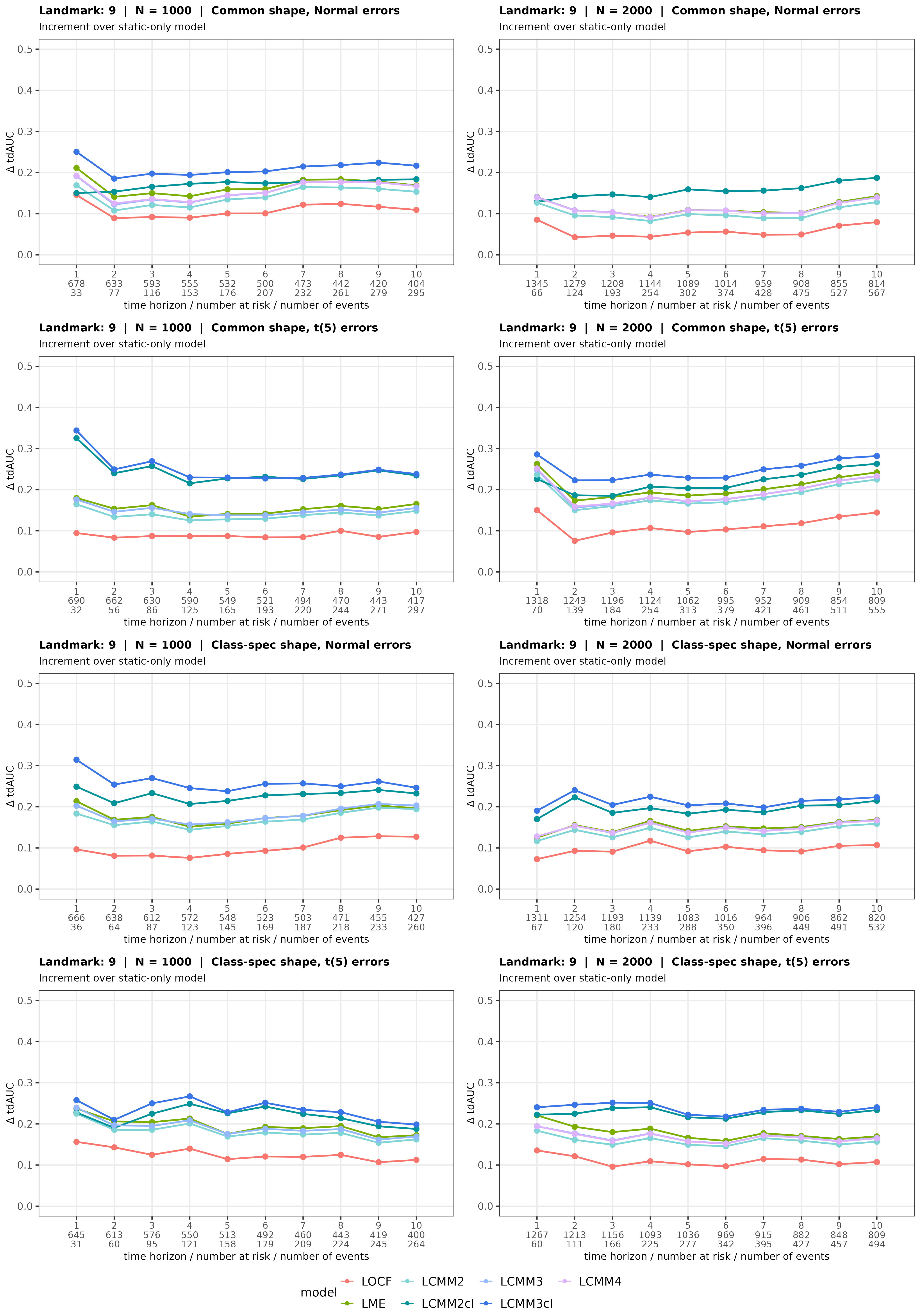}
  \caption{Cross-validated tdAUC increment over the static-only model at landmark 9, for the eight simulation scenarios, for the eight simulation scenarios. Lines show the increment in time-dependent AUC relative to the static-only Cox model, pooled across five cross-validation folds. Only models that converged in all five folds are shown. x-axis labels give the time horizon, number at risk, and number of events. Results are not shown for the models LCMM4 (N=1000, class-specific shape, t5 errors), LCMM4cl (N=1000, class-specific shape, normal errors) and LCMM4 (N=1000, class-specific shape, normal errors) as the optimiser failed to converge in one of the cross-validation training folds. }
  \label{fig:auct_lm9}
\end{figure}

\begin{figure}[H]
  \centering
  \includegraphics[width=\linewidth]{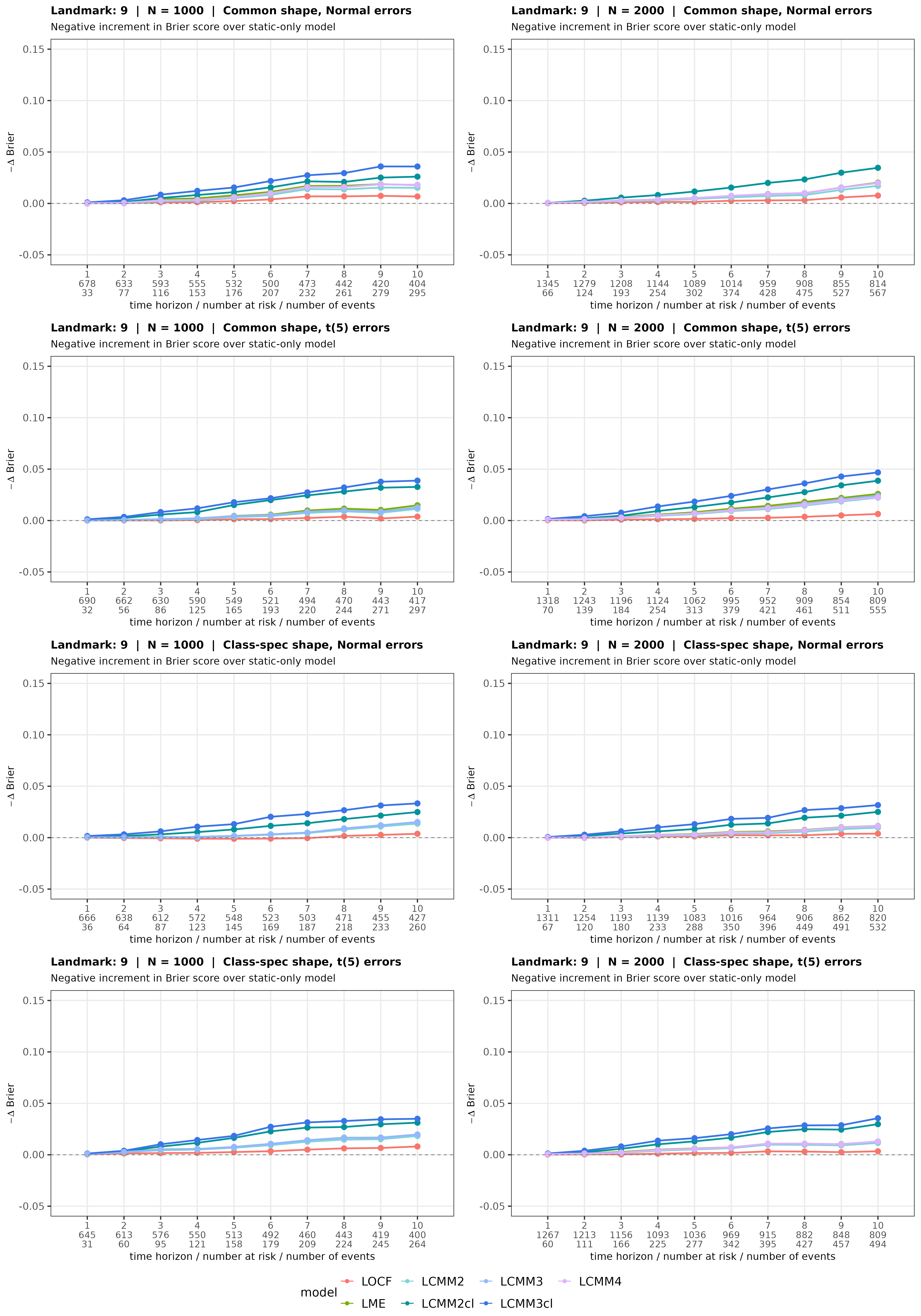}
  \caption{Negative increment in cross-validated Brier score over the static-only
           model at landmark 9, for the eight simulation scenarios. Positive values indicate improvement over the
           static-only Cox model.
           Only models that converged in all five folds are shown. Results are not shown for the models LCMM4 (N=1000, class-specific shape, t5 errors), LCMM4cl (N=1000, class-specific shape, normal errors) and LCMM4 (N=1000, class-specific shape, normal errors) as the optimiser failed to converge in one of the cross-validation training folds. }
  \label{fig:brier_lm9}
\end{figure}

\begin{figure}[H]
  \centering
  \includegraphics[width=\linewidth]{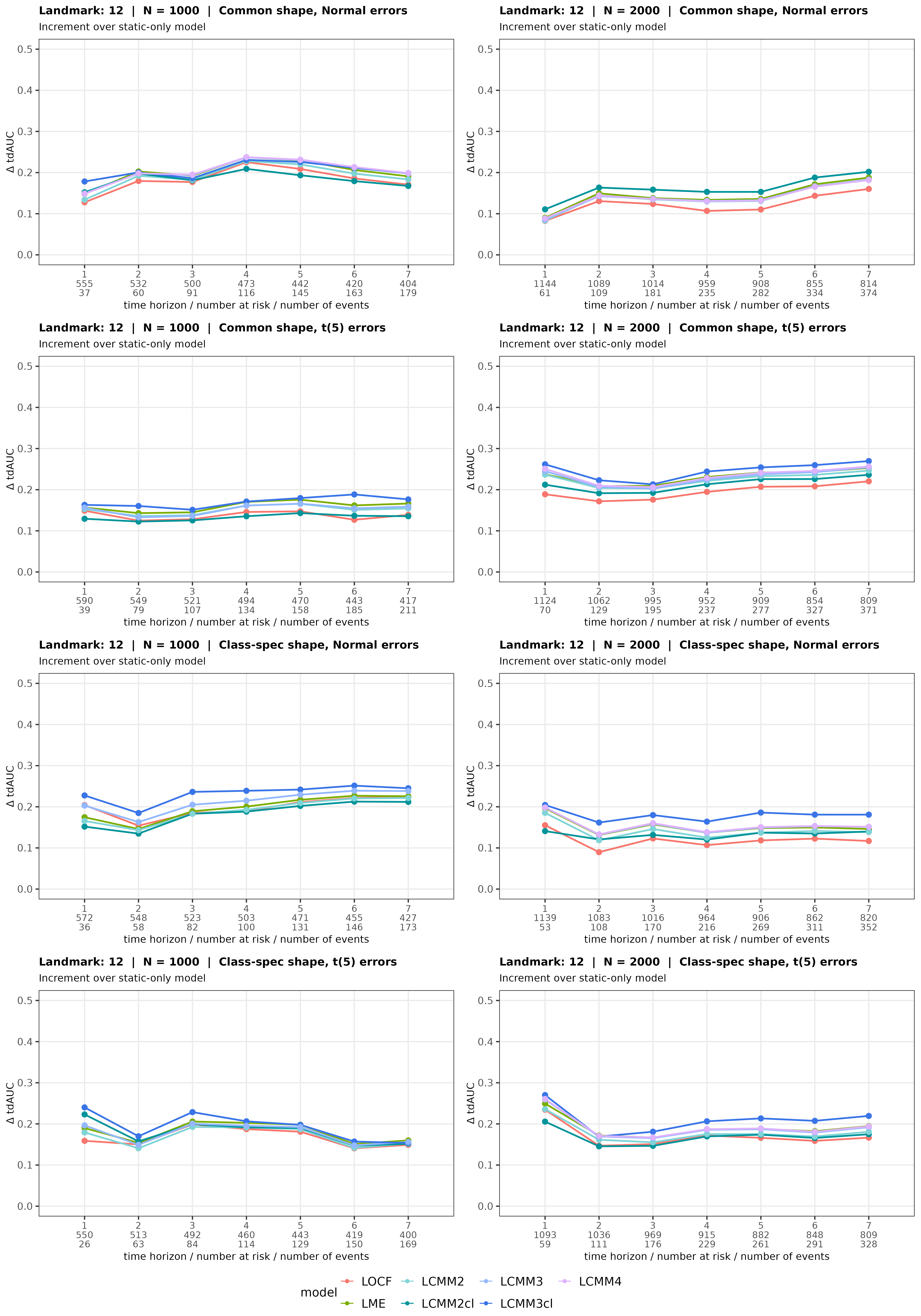}
  \caption{Cross-validated tdAUC increment over the static-only model at landmark 12, for the eight simulation scenarios. Lines show the increment in time-dependent AUC relative to the static-only Cox model, pooled across five cross-validation folds. Only models that converged in all five folds are shown. x-axis labels give the time horizon, number at risk, and number of events. Results are not shown for the models LCMM4 (N=1000, class-specific Weibull shape, t5 errors), LCMM4cl (N=1000, class-specific Weibull shape, normal errors) and LCMM4 (N=1000, class-specific Weibull shape, normal errors) as the optimiser failed to converge in one of the cross-validation training folds. }
  \label{fig:auct_lm12}
\end{figure}

\begin{figure}[H]
  \centering
  \includegraphics[width=\linewidth]{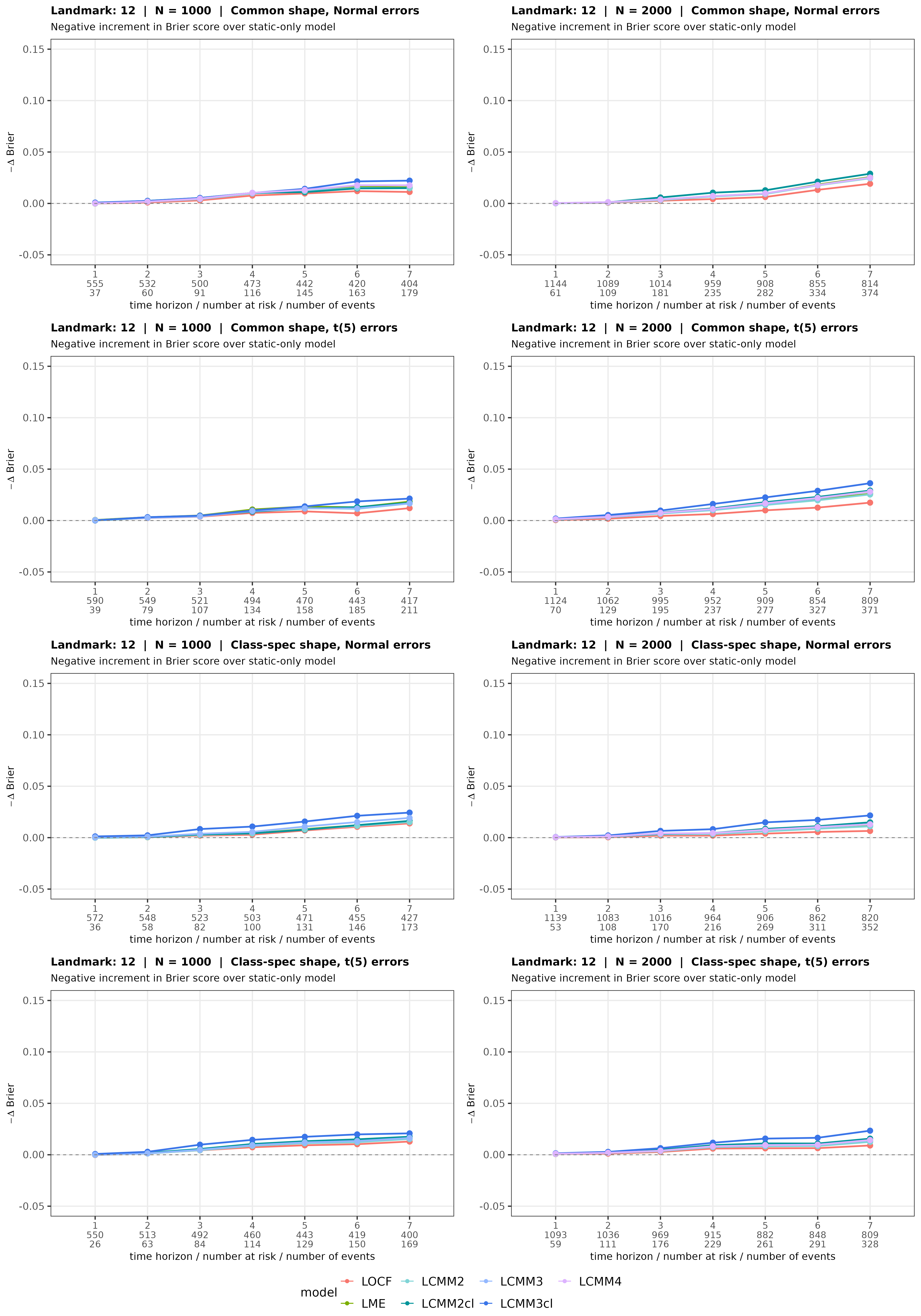}
  \caption{Negative increment in cross-validated Brier score over the static-only
           model at landmark 12, for the eight simulation scenarios. Positive values indicate improvement over the
           static-only Cox model.
           Only models that converged in all five folds are shown. Results are not shown for the models LCMM4 (N=1000, class-specific Weibull shape, t5 errors), LCMM4cl (N=1000, class-specific Weibull shape, normal errors) and LCMM4 (N=1000, class-specific Weibull shape, normal errors) as the optimiser failed to converge in one of the cross-validation training folds. }
  \label{fig:brier_lm12}
\end{figure}

\begin{landscape}
  \begin{table}[ht]
\footnotesize
\caption{Cross-validated tdAUC (95\% bootstrap CI): class-specific Weibull shape, $t(5)$ errors. Results for LCMM4 with $N=1000$ are omitted, as the
optimiser failed to converge in one cross-validation fold. }
\label{truncated-tablesauctClassSpect5tex}
\centering
\adjustbox{max width=\linewidth}{
\begin{tabular}[t]{cccrrrrrrrrrrrr}
\toprule
N & landmark & horizon & Static-only & LOCF & LME & LCMM2 & LCMM3 & LCMM4 & LCMM2cl & LCMM3cl & JLCM1 & JLCM2 & JLCM3 & JLCM4
\\
\midrule
1000 & 9 & 3 & 0.473 & 0.598 & 0.677 & 0.659 & 0.668 & -- & \textbf{0.698} & \textbf{0.723} & 0.477 & \textbf{0.684} & \textbf{0.739} & 0.423 \\
 &  &  & (0.411, 0.529) & (0.535, 0.663) & (0.626, 0.729) & (0.605, 0.710) & (0.615, 0.724) &  & \textbf{(0.646, 0.750)} & \textbf{(0.670, 0.775)} & (0.415, 0.542) & \textbf{(0.622, 0.745)} & \textbf{(0.689, 0.793)} & (0.371, 0.479) \\
1000 & 9 & 6 & 0.459 & 0.580 & 0.652 & 0.638 & 0.647 & -- & \textbf{0.701} & \textbf{0.711} & 0.484 & \textbf{0.659} & \textbf{0.718} & 0.414 \\
 &  &  & (0.408, 0.503) & (0.532, 0.627) & (0.608, 0.695) & (0.590, 0.684) & (0.602, 0.693) &  & \textbf{(0.662, 0.744)} & \textbf{(0.663, 0.757)} & (0.435, 0.533) & \textbf{(0.608, 0.713)} & \textbf{(0.674, 0.764)} & (0.371, 0.458) \\
1000 & 9 & 9 & 0.494 & 0.601 & 0.662 & 0.648 & 0.656 & -- & \textbf{0.688} & \textbf{0.699} & 0.488 & 0.641 & \textbf{0.701} & 0.387 \\
 &  &  & (0.449, 0.540) & (0.557, 0.646) & (0.620, 0.703) & (0.602, 0.691) & (0.613, 0.700) &  & \textbf{(0.648, 0.732)} & \textbf{(0.657, 0.742)} & (0.444, 0.533) & (0.595, 0.689) & \textbf{(0.659, 0.742)} & (0.344, 0.434) \\
1000 & 12 & 3 & 0.445 & 0.644 & 0.650 & 0.637 & 0.645 & -- & 0.644 & \textbf{0.673} & 0.499 & 0.616 & \textbf{0.681} & 0.517 \\
\bottomrule
\end{tabular}
}
\end{table}

  \end{landscape}
\begin{landscape}

  \begin{table}[ht]
\footnotesize
\caption{Cross-validated Brier score (95\% bootstrap CI): class-specific Weibull shape, $t(5)$ errors. Results for LCMM4 with $N=1000$ are omitted, as the optimiser failed to converge in one cross-validation fold.}
\label{truncated-tablesbrierClassSpect5tex}
\centering
\adjustbox{max width=\linewidth}{
\begin{tabular}[t]{cccrrrrrrrrrrrr}
\toprule
N & landmark & horizon & Static-only & LOCF & LME & LCMM2 & LCMM3 & LCMM4 & LCMM2cl & LCMM3cl & JLCM1 & JLCM2 & JLCM3 & JLCM4
\\
\midrule
1000 & 9 & 3 & 0.121 & 0.120 & 0.116 & 0.117 & 0.116 & -- & \textbf{0.113} & \textbf{0.111} & 0.224 & 0.206 & 0.199 & 0.465 \\
 &  &  & (0.104, 0.141) & (0.101, 0.138) & (0.098, 0.135) & (0.099, 0.135) & (0.099, 0.135) &  & \textbf{(0.096, 0.132)} & \textbf{(0.096, 0.128)} & (0.222, 0.226) & (0.195, 0.218) & (0.186, 0.211) & (0.441, 0.488) \\
1000 & 9 & 6 & 0.196 & 0.193 & 0.186 & 0.187 & \textbf{0.186} & -- & \textbf{0.174} & \textbf{0.169} & 0.273 & 0.241 & 0.232 & 0.481 \\
 &  &  & (0.182, 0.212) & (0.178, 0.209) & (0.172, 0.202) & (0.172, 0.202) & \textbf{(0.170, 0.201)} &  & \textbf{(0.160, 0.188)} & \textbf{(0.154, 0.183)} & (0.270, 0.276) & (0.231, 0.251) & (0.218, 0.244) & (0.455, 0.505) \\
1000 & 9 & 9 & 0.232 & 0.226 & 0.216 & 0.217 & \textbf{0.216} & -- & \textbf{0.203} & \textbf{0.198} & 0.295 & 0.265 & 0.251 & 0.489 \\
 &  &  & (0.224, 0.242) & (0.215, 0.237) & (0.204, 0.228) & (0.205, 0.230) & \textbf{(0.203, 0.227)} &  & \textbf{(0.190, 0.215)} & \textbf{(0.184, 0.213)} & (0.286, 0.303) & (0.253, 0.276) & (0.237, 0.264) & (0.461, 0.517) \\
1000 & 12 & 3 & 0.125 & 0.121 & 0.120 & 0.121 & 0.120 & -- & \textbf{0.120} & \textbf{0.115} & 0.292 & 0.285 & 0.290 & 0.512 \\
\bottomrule
\end{tabular}
}
\end{table}

\end{landscape}

\begin{landscape}
  \begin{table}[ht]
\footnotesize
\caption{Cross-validated tdAUC (95\% bootstrap CI): common Weibull shape, $t(5)$ errors. Results for LCMM4 with $N=1000$ are omitted, as the optimiser failed to converge in one cross-validation fold.}
\label{truncated-tablesauctCommonShapet5tex}
\centering
\adjustbox{max width=\linewidth}{
\begin{tabular}[t]{cccrrrrrrrrrrrr}
\toprule
N & landmark & horizon & Static-only & LOCF & LME & LCMM2 & LCMM3 & LCMM4 & LCMM2cl & LCMM3cl & JLCM1 & JLCM2 & JLCM3 & JLCM4
\\
\midrule
1000 & 9 & 3 & 0.436 & 0.523 & 0.598 & 0.576 & 0.592 & -- & \textbf{0.693} & \textbf{0.705} & 0.466 & \textbf{0.705} & 0.552 & 0.490 \\
 &  &  & (0.376, 0.498) & (0.461, 0.588) & (0.541, 0.651) & (0.522, 0.634) & (0.538, 0.647) &  & \textbf{(0.629, 0.750)} & \textbf{(0.638, 0.765)} & (0.403, 0.522) & \textbf{(0.637, 0.765)} & (0.488, 0.607) & (0.438, 0.542) \\
1000 & 9 & 6 & 0.460 & 0.545 & 0.602 & 0.590 & 0.598 & -- & \textbf{0.692} & \textbf{0.688} & 0.500 & \textbf{0.675} & 0.533 & 0.464 \\
 &  &  & (0.416, 0.502) & (0.501, 0.587) & (0.561, 0.647) & (0.550, 0.630) & (0.556, 0.643) &  & \textbf{(0.646, 0.737)} & \textbf{(0.640, 0.733)} & (0.455, 0.547) & \textbf{(0.626, 0.716)} & (0.487, 0.580) & (0.421, 0.508) \\
1000 & 9 & 9 & 0.458 & 0.544 & 0.611 & 0.596 & 0.602 & -- & \textbf{0.705} & \textbf{0.707} & 0.490 & \textbf{0.682} & 0.530 & 0.465 \\
 &  &  & (0.416, 0.498) & (0.501, 0.584) & (0.570, 0.654) & (0.553, 0.636) & (0.561, 0.643) &  & \textbf{(0.667, 0.745)} & \textbf{(0.662, 0.746)} & (0.447, 0.536) & \textbf{(0.640, 0.719)} & (0.488, 0.572) & (0.425, 0.505) \\
1000 & 12 & 3 & 0.485 & 0.613 & 0.630 & 0.623 & 0.621 & -- & 0.610 & \textbf{0.636} & 0.521 & \textbf{0.638} & 0.612 & 0.536 \\
\bottomrule
\end{tabular}
}
\end{table}

\end{landscape}

\begin{landscape}
  \begin{table}[ht]
\footnotesize
\caption{Cross-validated Brier score (95\% bootstrap CI): common Weibull shape, $t(5)$ errors. Results for LCMM4 with $N=1000$ are omitted, as the optimiser failed to converge in one cross-validation fold.}
\label{truncated-tablesbrierCommonShapet5tex}
\centering
\adjustbox{max width=\linewidth}{
\begin{tabular}[t]{cccrrrrrrrrrrrr}
\toprule
N & landmark & horizon & Static-only & LOCF & LME & LCMM2 & LCMM3 & LCMM4 & LCMM2cl & LCMM3cl & JLCM1 & JLCM2 & JLCM3 & JLCM4
\\
\midrule
1000 & 9 & 3 & 0.106 & 0.105 & 0.104 & 0.105 & 0.104 & -- & \textbf{0.100} & \textbf{0.097} & 0.241 & 0.218 & 0.335 & 0.432 \\
 &  &  & (0.087, 0.124) & (0.088, 0.125) & (0.087, 0.122) & (0.089, 0.122) & (0.087, 0.124) &  & \textbf{(0.083, 0.116)} & \textbf{(0.083, 0.115)} & (0.239, 0.243) & (0.206, 0.230) & (0.319, 0.351) & (0.412, 0.454) \\
1000 & 9 & 6 & 0.198 & 0.196 & 0.192 & 0.194 & 0.193 & -- & \textbf{0.178} & \textbf{0.176} & 0.293 & 0.255 & 0.372 & 0.449 \\
 &  &  & (0.183, 0.212) & (0.181, 0.213) & (0.176, 0.206) & (0.179, 0.208) & (0.178, 0.207) &  & \textbf{(0.163, 0.192)} & \textbf{(0.161, 0.192)} & (0.287, 0.298) & (0.242, 0.268) & (0.354, 0.390) & (0.427, 0.471) \\
1000 & 9 & 9 & 0.237 & 0.235 & 0.227 & 0.230 & 0.229 & -- & \textbf{0.205} & \textbf{0.200} & 0.313 & 0.264 & 0.391 & 0.445 \\
 &  &  & (0.229, 0.246) & (0.226, 0.245) & (0.216, 0.236) & (0.220, 0.240) & (0.218, 0.238) &  & \textbf{(0.193, 0.218)} & \textbf{(0.186, 0.215)} & (0.303, 0.324) & (0.251, 0.279) & (0.369, 0.414) & (0.421, 0.471) \\
1000 & 12 & 3 & 0.142 & 0.138 & 0.137 & 0.137 & 0.138 & -- & 0.137 & 0.137 & 0.328 & 0.311 & 0.421 & 0.465 \\
\bottomrule
\end{tabular}
}
\end{table}

\end{landscape}

\begin{landscape}
  \begin{table}[ht]
\footnotesize
\caption{Cross-validated tdAUC (95\% bootstrap CI): class-specific Weibull shape, normal errors. Results for LCMM4 with $N=1000$ are omitted, as the optimiser failed to converge in one cross-validation fold.}
\label{truncated-tablesauctClassSpecNormaltex}
\centering
\adjustbox{max width=\linewidth}{
\begin{tabular}[t]{cccrrrrrrrrrrrr}
\toprule
N & landmark & horizon & Static-only & LOCF & LME & LCMM2 & LCMM3 & LCMM4 & LCMM2cl & LCMM3cl & JLCM1 & JLCM2 & JLCM3 & JLCM4
\\
\midrule
1000 & 9 & 3 & 0.383 & 0.464 & 0.558 & 0.547 & 0.554 & -- & \textbf{0.616} & \textbf{0.653} & 0.432 & \textbf{0.672} & \textbf{0.667} & 0.425 \\
 &  &  & (0.326, 0.440) & (0.408, 0.528) & (0.500, 0.617) & (0.493, 0.605) & (0.494, 0.612) &  & \textbf{(0.558, 0.678)} & \textbf{(0.589, 0.714)} & (0.363, 0.503) & \textbf{(0.609, 0.731)} & \textbf{(0.600, 0.724)} & (0.370, 0.479) \\
1000 & 9 & 6 & 0.416 & 0.509 & 0.589 & 0.580 & 0.588 & -- & \textbf{0.644} & \textbf{0.672} & 0.428 & \textbf{0.683} & \textbf{0.699} & 0.426 \\
 &  &  & (0.367, 0.462) & (0.459, 0.559) & (0.541, 0.636) & (0.533, 0.625) & (0.537, 0.635) &  & \textbf{(0.592, 0.690)} & \textbf{(0.624, 0.726)} & (0.378, 0.476) & \textbf{(0.632, 0.730)} & \textbf{(0.650, 0.742)} & (0.381, 0.474) \\
1000 & 9 & 9 & 0.425 & 0.554 & 0.628 & 0.623 & \textbf{0.632} & -- & \textbf{0.666} & \textbf{0.687} & 0.436 & \textbf{0.677} & \textbf{0.709} & 0.401 \\
 &  &  & (0.378, 0.467) & (0.506, 0.603) & (0.580, 0.671) & (0.581, 0.665) & \textbf{(0.585, 0.672)} &  & \textbf{(0.621, 0.707)} & \textbf{(0.643, 0.729)} & (0.390, 0.477) & \textbf{(0.635, 0.720)} & \textbf{(0.668, 0.752)} & (0.361, 0.445) \\
1000 & 12 & 3 & 0.430 & 0.615 & 0.619 & 0.613 & \textbf{0.635} & -- & 0.613 & \textbf{0.666} & 0.432 & \textbf{0.682} & \textbf{0.710} & 0.579 \\
\bottomrule
\end{tabular}
}
\end{table}

\end{landscape}

\begin{landscape}
  \begin{table}[ht]
\footnotesize
\caption{Cross-validated Brier score (95\% bootstrap CI): class-specific Weibull shape, normal errors. Results for LCMM4 with $N=1000$ are omitted, as the optimiser failed to converge in one cross-validation fold.}
\label{truncated-tablesbrierClassSpecNormaltex}
\centering
\adjustbox{max width=\linewidth}{
\begin{tabular}[t]{cccrrrrrrrrrrrr}
\toprule
N & landmark & horizon & Static-only & LOCF & LME & LCMM2 & LCMM3 & LCMM4 & LCMM2cl & LCMM3cl & JLCM1 & JLCM2 & JLCM3 & JLCM4
\\
\midrule
1000 & 9 & 3 & 0.110 & 0.110 & 0.109 & 0.109 & 0.109 & -- & \textbf{0.106} & \textbf{0.104} & 0.208 & 0.192 & 0.195 & 0.463 \\
 &  &  & (0.091, 0.130) & (0.093, 0.128) & (0.092, 0.128) & (0.092, 0.128) & (0.091, 0.128) &  & \textbf{(0.090, 0.124)} & \textbf{(0.086, 0.121)} & (0.205, 0.212) & (0.180, 0.204) & (0.183, 0.208) & (0.440, 0.488) \\
1000 & 9 & 6 & 0.185 & 0.186 & 0.182 & 0.182 & 0.182 & -- & \textbf{0.174} & \textbf{0.165} & 0.262 & 0.230 & 0.226 & 0.477 \\
 &  &  & (0.168, 0.202) & (0.170, 0.203) & (0.165, 0.199) & (0.167, 0.200) & (0.165, 0.197) &  & \textbf{(0.158, 0.189)} & \textbf{(0.147, 0.181)} & (0.261, 0.264) & (0.219, 0.241) & (0.213, 0.240) & (0.454, 0.500) \\
1000 & 9 & 9 & 0.225 & 0.222 & 0.214 & 0.214 & \textbf{0.213} & -- & \textbf{0.204} & \textbf{0.194} & 0.290 & 0.255 & 0.242 & 0.485 \\
 &  &  & (0.213, 0.237) & (0.211, 0.235) & (0.201, 0.227) & (0.202, 0.227) & \textbf{(0.201, 0.225)} &  & \textbf{(0.190, 0.217)} & \textbf{(0.178, 0.208)} & (0.284, 0.296) & (0.244, 0.267) & (0.227, 0.257) & (0.461, 0.510) \\
1000 & 12 & 3 & 0.117 & 0.115 & 0.114 & 0.114 & \textbf{0.113} & -- & \textbf{0.114} & \textbf{0.109} & 0.277 & 0.267 & 0.276 & 0.479 \\
\bottomrule
\end{tabular}
}
\end{table}

\end{landscape}

\begin{landscape}
  \begin{table}[ht]
\footnotesize
\caption{Cross-validated tdAUC (95\% bootstrap CI): common Weibull shape, normal errors. Results for JLCM3 with $N=1000$ and LCMM3cl with $N=2000$ are omitted, as the optimiser failed to converge in one cross-validation fold.}
\label{truncated-tablesauctCommonShapeNormaltex}
\centering
\adjustbox{max width=\linewidth}{
\begin{tabular}[t]{cccrrrrrrrrrrrr}
\toprule
N & landmark & horizon & Static-only & LOCF & LME & LCMM2 & LCMM3 & LCMM4 & LCMM2cl & LCMM3cl & JLCM1 & JLCM2 & JLCM3 & JLCM4
\\
\midrule
1000 & 9 & 3 & 0.473 & 0.565 & 0.623 & 0.595 & 0.608 & 0.609 & \textbf{0.639} & \textbf{0.671} & 0.498 & 0.618 & -- & 0.440 \\
 &  &  & (0.416, 0.528) & (0.510, 0.618) & (0.569, 0.676) & (0.544, 0.646) & (0.551, 0.659) & (0.559, 0.661) & \textbf{(0.587, 0.693)} & \textbf{(0.611, 0.724)} & (0.444, 0.552) & (0.554, 0.670) &  & (0.390, 0.491) \\
1000 & 9 & 6 & 0.483 & 0.584 & 0.643 & 0.622 & 0.633 & 0.634 & \textbf{0.657} & \textbf{0.685} & 0.506 & 0.614 & -- & 0.435 \\
 &  &  & (0.433, 0.529) & (0.539, 0.626) & (0.599, 0.681) & (0.579, 0.665) & (0.587, 0.677) & (0.590, 0.674) & \textbf{(0.609, 0.699)} & \textbf{(0.642, 0.731)} & (0.459, 0.551) & (0.570, 0.659) &  & (0.389, 0.481) \\
1000 & 9 & 9 & 0.484 & 0.601 & 0.662 & 0.644 & 0.660 & 0.660 & \textbf{0.666} & \textbf{0.708} & 0.491 & 0.589 & -- & 0.427 \\
 &  &  & (0.440, 0.527) & (0.561, 0.643) & (0.618, 0.700) & (0.603, 0.686) & (0.616, 0.701) & (0.620, 0.700) & \textbf{(0.625, 0.707)} & \textbf{(0.666, 0.749)} & (0.447, 0.534) & (0.546, 0.627) &  & (0.385, 0.469) \\
1000 & 12 & 3 & 0.449 & 0.626 & 0.641 & 0.631 & \textbf{0.643} & \textbf{0.644} & 0.630 & 0.635 & 0.516 & 0.623 & -- & 0.583 \\
\bottomrule
\end{tabular}
}
\end{table}

\end{landscape}

\begin{landscape}
  \begin{table}[ht]
\footnotesize
\caption{Cross-validated Brier score (95\% bootstrap CI): common Weibull shape, normal errors. Results for JLCM3 with $N=1000$ and LCMM3cl with $N=2000$ are omitted, as the optimiser failed to converge in one cross-validation fold.}
\label{truncated-tablesbrierCommonShapeNormaltex}
\centering
\adjustbox{max width=\linewidth}{
\begin{tabular}[t]{cccrrrrrrrrrrrr}
\toprule
N & landmark & horizon & Static-only & LOCF & LME & LCMM2 & LCMM3 & LCMM4 & LCMM2cl & LCMM3cl & JLCM1 & JLCM2 & JLCM3 & JLCM4
\\
\midrule
1000 & 9 & 3 & 0.137 & 0.136 & 0.133 & 0.135 & 0.134 & 0.134 & \textbf{0.132} & \textbf{0.128} & 0.242 & 0.238 & -- & 0.475 \\
 &  &  & (0.119, 0.154) & (0.118, 0.156) & (0.116, 0.152) & (0.117, 0.153) & (0.117, 0.151) & (0.117, 0.151) & \textbf{(0.114, 0.148)} & \textbf{(0.112, 0.145)} & (0.240, 0.243) & (0.226, 0.250) &  & (0.453, 0.499) \\
1000 & 9 & 6 & 0.207 & 0.203 & 0.196 & 0.199 & 0.197 & 0.197 & \textbf{0.192} & \textbf{0.186} & 0.288 & 0.299 & -- & 0.470 \\
 &  &  & (0.193, 0.221) & (0.189, 0.218) & (0.183, 0.211) & (0.185, 0.212) & (0.183, 0.212) & (0.184, 0.211) & \textbf{(0.178, 0.205)} & \textbf{(0.170, 0.200)} & (0.282, 0.293) & (0.285, 0.316) &  & (0.445, 0.496) \\
1000 & 9 & 9 & 0.240 & 0.232 & 0.221 & 0.224 & 0.221 & 0.221 & \textbf{0.215} & \textbf{0.204} & 0.306 & 0.340 & -- & 0.451 \\
 &  &  & (0.232, 0.247) & (0.224, 0.242) & (0.211, 0.232) & (0.214, 0.234) & (0.211, 0.232) & (0.211, 0.232) & \textbf{(0.203, 0.227)} & \textbf{(0.190, 0.218)} & (0.295, 0.317) & (0.319, 0.363) &  & (0.425, 0.480) \\
1000 & 12 & 3 & 0.131 & 0.128 & 0.126 & 0.126 & \textbf{0.126} & \textbf{0.126} & \textbf{0.125} & \textbf{0.125} & 0.326 & 0.403 & -- & 0.447 \\
\bottomrule
\end{tabular}
}
\end{table}

\end{landscape}

\subsubsection{Results: simulated data (supplementary analysis)} \label{sec:results-simulations-supplementary}

\begin{figure}[p]
  \centering
  \includegraphics[width=\linewidth]{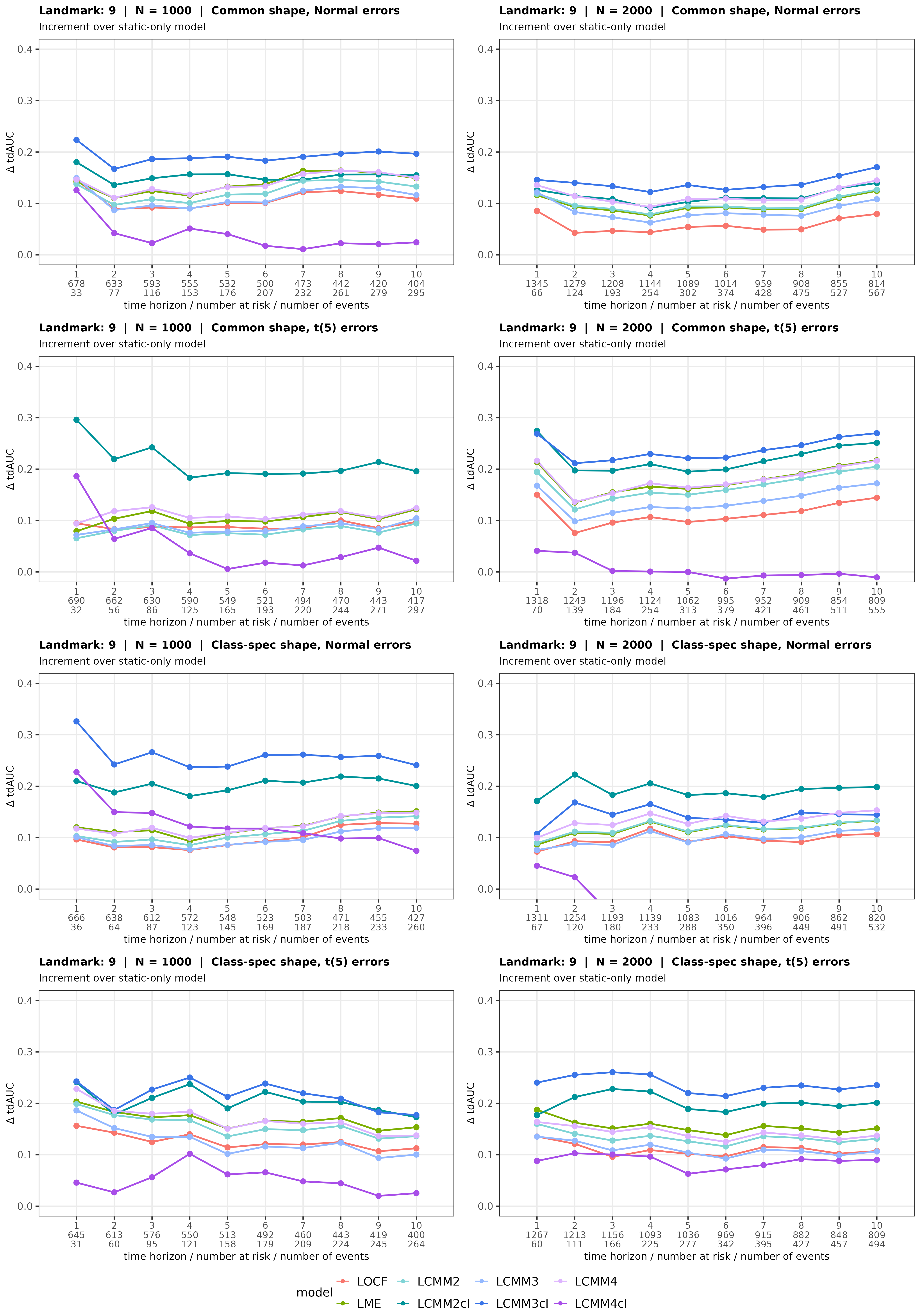}
  \caption{Cross-validated tdAUC increment over the static-only model at landmark 9, for the eight simulation scenarios.
           Lines show the increment in time-dependent AUC relative to the static-only
           Cox model, pooled across five cross-validation folds.
           Only models that converged in all five folds are shown.
           x-axis labels give the time horizon, number at risk, and number of events. Results are not shown for the models LCMM3cl (N=1000, common Weibull shape, $t_5$ errors) as the optimiser failed to converge in one of the cross-validation training folds.}
  \label{fig:supp-auct_lm9}
\end{figure}

\begin{figure}[p]
  \centering
  \includegraphics[width=\linewidth]{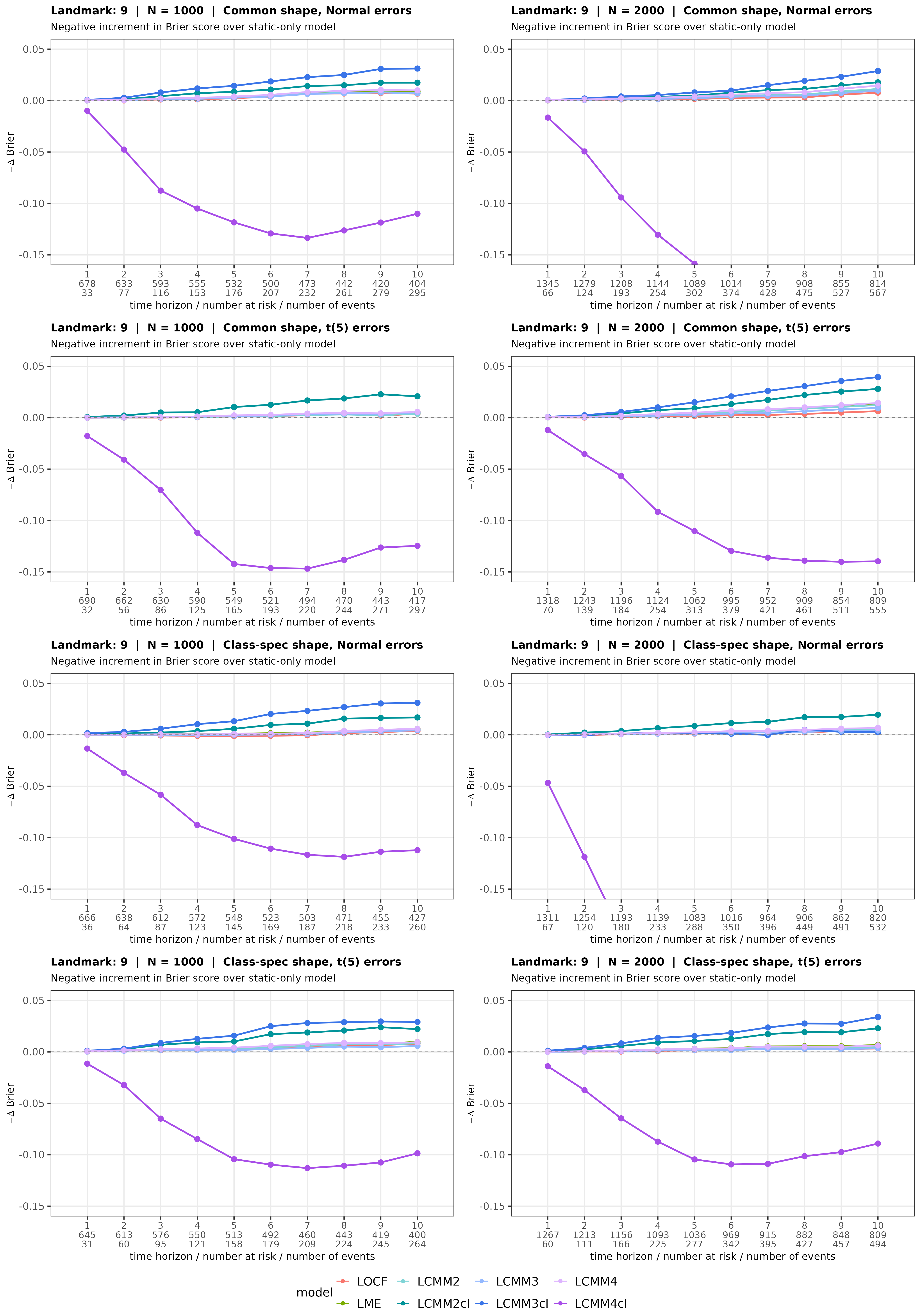}
  \caption{Negative increment in cross-validated Brier score over the static-only
           model at landmark 9, for the eight simulation scenarios. Positive values indicate improvement over the
           static-only Cox model.
           Only models that converged in all five folds are shown.}
  \label{fig:supp-brier_lm9}
\end{figure}

\begin{figure}[p]
  \centering
  \includegraphics[width=\linewidth]{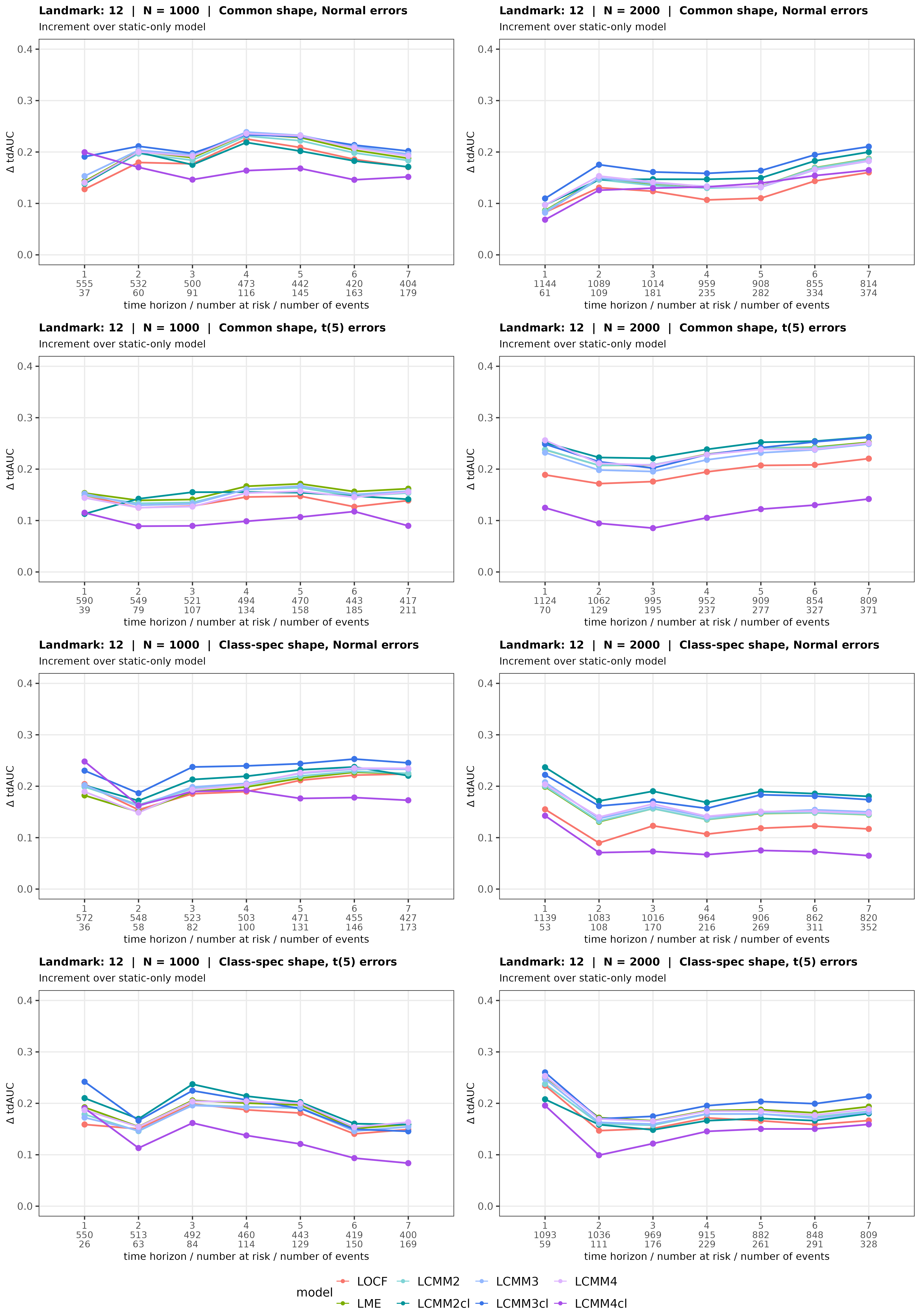}
  \caption{Cross-validated tdAUC increment over the static-only model at landmark 12, for the eight simulation scenarios.}
  \label{fig:supp-auct_lm12}
\end{figure}

\begin{figure}[p]
  \centering
  \includegraphics[width=\linewidth]{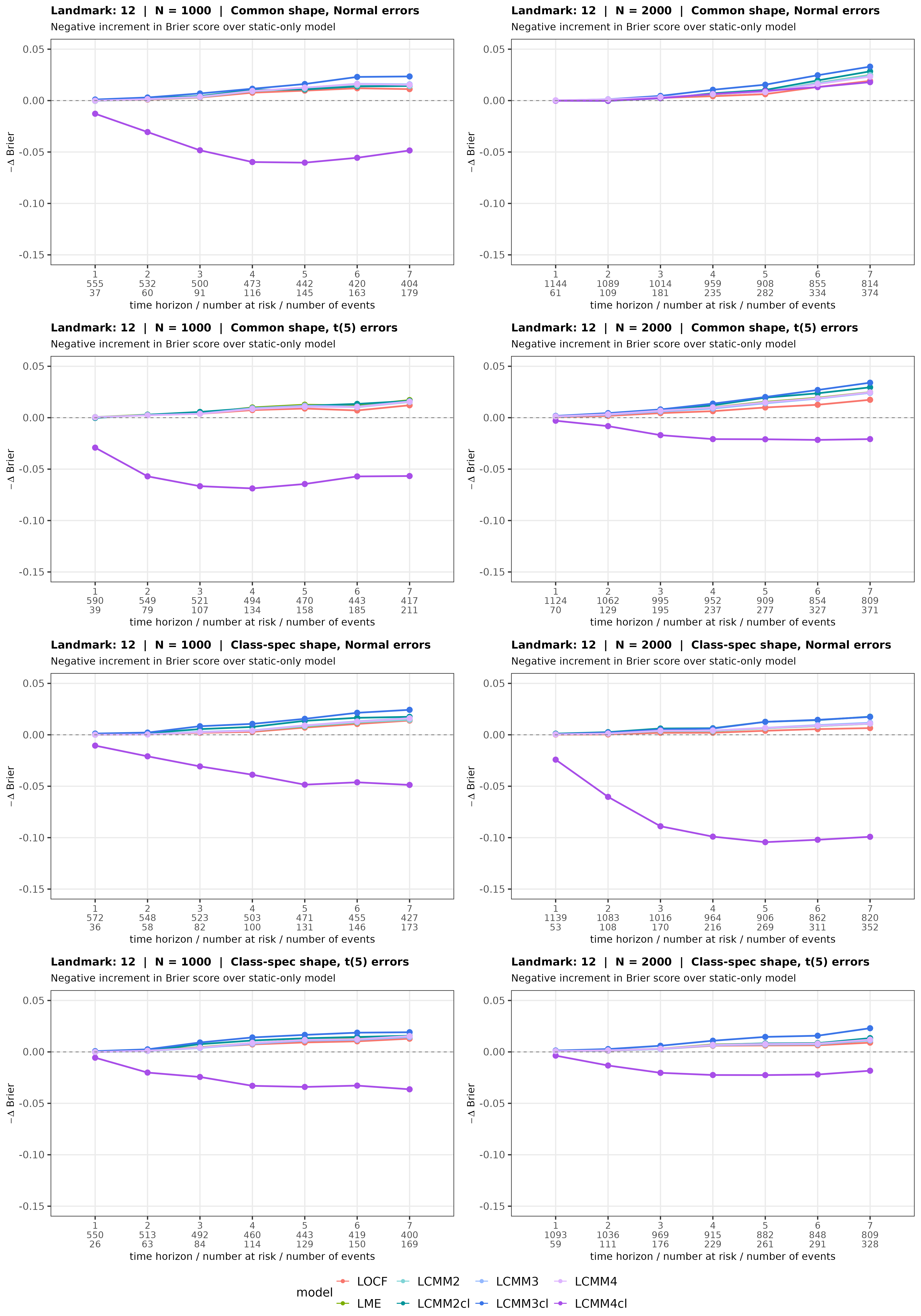}
  \caption{Negative increment in cross-validated Brier score over the static-only
           model at landmark 9, for the eight simulation scenarios.}
  \label{fig:supp-brier_lm12}
\end{figure}

\begin{landscape}
  \begin{table}[ht]
\footnotesize
\caption{Cross-validated tdAUC (95\% bootstrap CI): class-specific Weibull shape, $t(5)$ errors.}
\label{nontruncated-tablesauctClassSpect5tex}
\centering
\adjustbox{max width=\linewidth}{
\begin{tabular}[t]{cccrrrrrrrrrrrrr}
\toprule
N & landmark & horizon & Static-only & LOCF & LME & LCMM2 & LCMM3 & LCMM4 & LCMM2cl & LCMM3cl & LCMM4cl & JLCM1 & JLCM2 & JLCM3 & JLCM4
\\
\midrule
1000 & 9 & 3 & 0.473 & 0.598 & 0.646 & 0.641 & 0.608 & \textbf{0.653} & \textbf{0.683} & \textbf{0.700} & 0.529 & 0.477 & \textbf{0.684} & \textbf{0.739} & 0.423 \\
 &  &  & (0.416, 0.533) & (0.536, 0.666) & (0.594, 0.700) & (0.580, 0.695) & (0.545, 0.666) & \textbf{(0.598, 0.705)} & \textbf{(0.619, 0.741)} & \textbf{(0.641, 0.752)} & (0.482, 0.581) & (0.412, 0.538) & \textbf{(0.623, 0.745)} & \textbf{(0.684, 0.787)} & (0.369, 0.474) \\
1000 & 9 & 6 & 0.459 & 0.580 & 0.625 & 0.609 & 0.575 & 0.625 & \textbf{0.681} & \textbf{0.698} & 0.525 & 0.484 & \textbf{0.659} & \textbf{0.718} & 0.414 \\
 &  &  & (0.412, 0.505) & (0.532, 0.628) & (0.577, 0.673) & (0.563, 0.655) & (0.526, 0.622) & (0.578, 0.671) & \textbf{(0.630, 0.729)} & \textbf{(0.650, 0.744)} & (0.481, 0.569) & (0.434, 0.536) & \textbf{(0.608, 0.709)} & \textbf{(0.672, 0.761)} & (0.370, 0.460) \\
1000 & 9 & 9 & 0.494 & 0.601 & 0.641 & 0.626 & 0.588 & 0.631 & \textbf{0.681} & \textbf{0.677} & 0.514 & 0.488 & \textbf{0.641} & \textbf{0.701} & 0.387 \\
 &  &  & (0.452, 0.537) & (0.556, 0.649) & (0.596, 0.686) & (0.580, 0.672) & (0.544, 0.631) & (0.586, 0.674) & \textbf{(0.637, 0.725)} & \textbf{(0.630, 0.722)} & (0.470, 0.559) & (0.444, 0.537) & \textbf{(0.596, 0.685)} & \textbf{(0.659, 0.741)} & (0.343, 0.430) \\
1000 & 12 & 3 & 0.445 & 0.644 & 0.650 & 0.642 & 0.641 & 0.648 & \textbf{0.682} & \textbf{0.669} & 0.606 & 0.499 & 0.616 & \textbf{0.681} & 0.517 \\
\bottomrule
\end{tabular}
}
\end{table}

\end{landscape}

\begin{landscape}
  \begin{table}[ht]
\footnotesize
\caption{Cross-validated Brier score (95\% bootstrap CI): class-specific Weibull shape, $t(5)$ errors.}
\label{nontruncated-tablesbrierClassSpect5tex}
\centering
\adjustbox{max width=\linewidth}{
\begin{tabular}[t]{cccrrrrrrrrrrrrr}
\toprule
N & landmark & horizon & Static-only & LOCF & LME & LCMM2 & LCMM3 & LCMM4 & LCMM2cl & LCMM3cl & LCMM4cl & JLCM1 & JLCM2 & JLCM3 & JLCM4
\\
\midrule
1000 & 9 & 3 & 0.121 & 0.120 & 0.119 & 0.119 & 0.119 & \textbf{0.118} & \textbf{0.114} & \textbf{0.113} & 0.186 & 0.224 & 0.206 & 0.199 & 0.465 \\
 &  &  & (0.102, 0.141) & (0.101, 0.140) & (0.100, 0.138) & (0.100, 0.138) & (0.100, 0.140) & \textbf{(0.100, 0.139)} & \textbf{(0.097, 0.133)} & \textbf{(0.095, 0.131)} & (0.167, 0.206) & (0.222, 0.226) & (0.195, 0.217) & (0.187, 0.212) & (0.443, 0.489) \\
1000 & 9 & 6 & 0.196 & 0.193 & 0.191 & 0.192 & 0.194 & \textbf{0.190} & \textbf{0.179} & \textbf{0.171} & 0.306 & 0.273 & 0.241 & 0.232 & 0.481 \\
 &  &  & (0.180, 0.213) & (0.178, 0.209) & (0.174, 0.208) & (0.175, 0.208) & (0.177, 0.211) & \textbf{(0.173, 0.208)} & \textbf{(0.163, 0.196)} & \textbf{(0.156, 0.189)} & (0.282, 0.330) & (0.270, 0.276) & (0.231, 0.252) & (0.219, 0.244) & (0.456, 0.505) \\
1000 & 9 & 9 & 0.232 & 0.226 & 0.224 & 0.225 & 0.228 & \textbf{0.224} & \textbf{0.208} & \textbf{0.203} & 0.340 & 0.295 & 0.265 & 0.251 & 0.489 \\
 &  &  & (0.223, 0.243) & (0.215, 0.236) & (0.212, 0.237) & (0.213, 0.238) & (0.216, 0.240) & \textbf{(0.210, 0.237)} & \textbf{(0.195, 0.222)} & \textbf{(0.188, 0.219)} & (0.315, 0.365) & (0.287, 0.303) & (0.254, 0.276) & (0.238, 0.266) & (0.461, 0.519) \\
1000 & 12 & 3 & 0.125 & 0.121 & 0.121 & 0.121 & 0.121 & 0.121 & \textbf{0.118} & \textbf{0.116} & 0.150 & 0.292 & 0.285 & 0.290 & 0.512 \\
\bottomrule
\end{tabular}
}
\end{table}

\end{landscape}

\begin{landscape}
  \begin{table}[ht]
\footnotesize
\caption{Cross-validated tdAUC (95\% bootstrap CI): common Weibull shape, $t(5)$ errors. Results for LCMMcl with $N=1000$ are omitted, as the optimiser failed to converge in one cross-validation fold.}
\label{nontruncated-tablesauctCommonShapet5tex}
\centering
\adjustbox{max width=\linewidth}{
\begin{tabular}[t]{cccrrrrrrrrrrrrr}
\toprule
N & landmark & horizon & Static-only & LOCF & LME & LCMM2 & LCMM3 & LCMM4 & LCMM2cl & LCMM3cl & LCMM4cl & JLCM1 & JLCM2 & JLCM3 & JLCM4
\\
\midrule
1000 & 9 & 3 & 0.436 & 0.523 & 0.554 & 0.526 & 0.531 & \textbf{0.562} & \textbf{0.678} & -- & 0.521 & 0.466 & \textbf{0.705} & 0.552 & 0.490 \\
 &  &  & (0.375, 0.496) & (0.457, 0.587) & (0.492, 0.613) & (0.467, 0.586) & (0.472, 0.591) & \textbf{(0.503, 0.616)} & \textbf{(0.616, 0.734)} &  & (0.467, 0.575) & (0.407, 0.527) & \textbf{(0.641, 0.766)} & (0.490, 0.615) & (0.436, 0.543) \\
1000 & 9 & 6 & 0.460 & 0.545 & 0.559 & 0.533 & 0.540 & \textbf{0.563} & \textbf{0.651} & -- & 0.479 & 0.500 & \textbf{0.675} & 0.533 & 0.464 \\
 &  &  & (0.415, 0.504) & (0.498, 0.589) & (0.514, 0.603) & (0.491, 0.574) & (0.498, 0.585) & \textbf{(0.519, 0.606)} & \textbf{(0.605, 0.699)} &  & (0.437, 0.525) & (0.455, 0.547) & \textbf{(0.628, 0.719)} & (0.485, 0.577) & (0.421, 0.505) \\
1000 & 9 & 9 & 0.458 & 0.544 & 0.561 & 0.535 & 0.541 & \textbf{0.563} & \textbf{0.672} & -- & 0.506 & 0.490 & \textbf{0.682} & 0.530 & 0.465 \\
 &  &  & (0.416, 0.500) & (0.500, 0.587) & (0.517, 0.602) & (0.494, 0.574) & (0.499, 0.584) & \textbf{(0.523, 0.604)} & \textbf{(0.629, 0.715)} &  & (0.464, 0.551) & (0.446, 0.535) & \textbf{(0.639, 0.723)} & (0.491, 0.576) & (0.427, 0.505) \\
1000 & 12 & 3 & 0.485 & 0.613 & 0.626 & 0.620 & 0.617 & 0.612 & \textbf{0.640} & -- & 0.574 & 0.521 & \textbf{0.638} & 0.612 & 0.536 \\
\bottomrule
\end{tabular}
}
\end{table}

\end{landscape}

\begin{landscape}
  \begin{table}[ht]
\footnotesize
\caption{Cross-validated Brier score (95\% bootstrap CI): common Weibull shape, $t(5)$ errors. Results for LCMMcl with $N=1000$ are omitted, as the optimiser failed to converge in one cross-validation fold.}
\label{nontruncated-tablesbrierCommonShapet5tex}
\centering
\adjustbox{max width=\linewidth}{
\begin{tabular}[t]{cccrrrrrrrrrrrrr}
\toprule
N & landmark & horizon & Static-only & LOCF & LME & LCMM2 & LCMM3 & LCMM4 & LCMM2cl & LCMM3cl & LCMM4cl & JLCM1 & JLCM2 & JLCM3 & JLCM4
\\
\midrule
1000 & 9 & 3 & 0.106 & 0.105 & 0.105 & 0.105 & 0.105 & \textbf{0.105} & \textbf{0.100} & -- & 0.176 & 0.241 & 0.218 & 0.335 & 0.432 \\
 &  &  & (0.087, 0.123) & (0.089, 0.124) & (0.087, 0.122) & (0.087, 0.122) & (0.089, 0.123) & \textbf{(0.087, 0.123)} & \textbf{(0.085, 0.118)} &  & (0.159, 0.194) & (0.239, 0.243) & (0.206, 0.231) & (0.317, 0.354) & (0.412, 0.453) \\
1000 & 9 & 6 & 0.198 & 0.196 & 0.195 & 0.196 & 0.196 & \textbf{0.195} & \textbf{0.185} & -- & 0.344 & 0.293 & 0.255 & 0.372 & 0.449 \\
 &  &  & (0.183, 0.213) & (0.181, 0.211) & (0.179, 0.211) & (0.180, 0.211) & (0.181, 0.211) & \textbf{(0.179, 0.210)} & \textbf{(0.171, 0.201)} &  & (0.320, 0.367) & (0.287, 0.298) & (0.242, 0.268) & (0.355, 0.392) & (0.426, 0.471) \\
1000 & 9 & 9 & 0.237 & 0.235 & 0.233 & 0.235 & 0.234 & \textbf{0.233} & \textbf{0.214} & -- & 0.363 & 0.313 & 0.264 & 0.391 & 0.445 \\
 &  &  & (0.228, 0.245) & (0.226, 0.245) & (0.223, 0.243) & (0.224, 0.245) & (0.224, 0.243) & \textbf{(0.223, 0.243)} & \textbf{(0.202, 0.228)} &  & (0.337, 0.389) & (0.303, 0.324) & (0.251, 0.278) & (0.370, 0.415) & (0.420, 0.469) \\
1000 & 12 & 3 & 0.142 & 0.138 & 0.137 & 0.138 & 0.138 & 0.138 & \textbf{0.136} & -- & 0.208 & 0.328 & 0.311 & 0.421 & 0.465 \\
\bottomrule
\end{tabular}
}
\end{table}

\end{landscape}

\begin{landscape}
  \begin{table}[ht]
\footnotesize
\caption{Cross-validated tdAUC (95\% bootstrap CI): class-specific Weibull shape, normal errors.}
\label{nontruncated-tablesauctClassSpecNormaltex}
\centering
\adjustbox{max width=\linewidth}{
\begin{tabular}[t]{cccrrrrrrrrrrrrr}
\toprule
N & landmark & horizon & Static-only & LOCF & LME & LCMM2 & LCMM3 & LCMM4 & LCMM2cl & LCMM3cl & LCMM4cl & JLCM1 & JLCM2 & JLCM3 & JLCM4
\\
\midrule
1000 & 9 & 3 & 0.383 & 0.464 & 0.497 & 0.479 & 0.469 & \textbf{0.502} & \textbf{0.588} & \textbf{0.649} & \textbf{0.531} & 0.432 & \textbf{0.672} & \textbf{0.667} & 0.425 \\
 &  &  & (0.327, 0.439) & (0.402, 0.530) & (0.439, 0.555) & (0.421, 0.540) & (0.407, 0.527) & \textbf{(0.440, 0.563)} & \textbf{(0.520, 0.651)} & \textbf{(0.576, 0.711)} & \textbf{(0.474, 0.587)} & (0.370, 0.499) & \textbf{(0.614, 0.735)} & \textbf{(0.601, 0.733)} & (0.370, 0.480) \\
1000 & 9 & 6 & 0.416 & 0.509 & 0.534 & 0.523 & 0.508 & \textbf{0.535} & \textbf{0.627} & \textbf{0.677} & 0.534 & 0.428 & \textbf{0.683} & \textbf{0.699} & 0.426 \\
 &  &  & (0.369, 0.461) & (0.462, 0.557) & (0.487, 0.579) & (0.476, 0.571) & (0.460, 0.554) & \textbf{(0.487, 0.583)} & \textbf{(0.578, 0.675)} & \textbf{(0.628, 0.725)} & (0.489, 0.576) & (0.382, 0.477) & \textbf{(0.637, 0.732)} & \textbf{(0.655, 0.741)} & (0.382, 0.469) \\
1000 & 9 & 9 & 0.425 & 0.554 & 0.574 & 0.564 & 0.544 & 0.573 & \textbf{0.640} & \textbf{0.684} & 0.524 & 0.436 & \textbf{0.677} & \textbf{0.709} & 0.401 \\
 &  &  & (0.384, 0.464) & (0.511, 0.600) & (0.531, 0.616) & (0.521, 0.610) & (0.502, 0.590) & (0.527, 0.621) & \textbf{(0.593, 0.686)} & \textbf{(0.643, 0.732)} & (0.479, 0.567) & (0.391, 0.479) & \textbf{(0.636, 0.720)} & \textbf{(0.670, 0.749)} & (0.362, 0.443) \\
1000 & 12 & 3 & 0.430 & 0.615 & 0.620 & \textbf{0.625} & \textbf{0.629} & \textbf{0.623} & \textbf{0.643} & \textbf{0.668} & \textbf{0.620} & 0.432 & \textbf{0.682} & \textbf{0.710} & 0.579 \\
\bottomrule
\end{tabular}
}
\end{table}

\end{landscape}

\begin{landscape}
  \begin{table}[ht]
\footnotesize
\caption{Cross-validated Brier score (95\% bootstrap CI): class-specific Weibull shape, normal errors.}
\label{nontruncated-tablesbrierClassSpecNormaltex}
\centering
\adjustbox{max width=\linewidth}{
\begin{tabular}[t]{cccrrrrrrrrrrrrr}
\toprule
N & landmark & horizon & Static-only & LOCF & LME & LCMM2 & LCMM3 & LCMM4 & LCMM2cl & LCMM3cl & LCMM4cl & JLCM1 & JLCM2 & JLCM3 & JLCM4
\\
\midrule
1000 & 9 & 3 & 0.110 & 0.110 & 0.109 & 0.109 & 0.110 & 0.109 & \textbf{0.108} & \textbf{0.104} & 0.168 & 0.208 & 0.192 & 0.195 & 0.463 \\
 &  &  & (0.091, 0.128) & (0.092, 0.130) & (0.090, 0.128) & (0.091, 0.128) & (0.090, 0.127) & (0.090, 0.130) & \textbf{(0.090, 0.126)} & \textbf{(0.086, 0.122)} & (0.151, 0.186) & (0.205, 0.211) & (0.181, 0.204) & (0.181, 0.208) & (0.441, 0.486) \\
1000 & 9 & 6 & 0.185 & 0.186 & 0.184 & 0.184 & 0.185 & 0.184 & \textbf{0.176} & \textbf{0.165} & 0.296 & 0.262 & 0.230 & 0.226 & 0.477 \\
 &  &  & (0.168, 0.201) & (0.169, 0.202) & (0.165, 0.201) & (0.167, 0.200) & (0.167, 0.201) & (0.167, 0.202) & \textbf{(0.159, 0.192)} & \textbf{(0.149, 0.182)} & (0.274, 0.319) & (0.261, 0.264) & (0.219, 0.241) & (0.212, 0.239) & (0.454, 0.501) \\
1000 & 9 & 9 & 0.225 & 0.222 & 0.220 & 0.221 & 0.222 & \textbf{0.220} & \textbf{0.209} & \textbf{0.195} & 0.339 & 0.290 & 0.255 & 0.242 & 0.485 \\
 &  &  & (0.213, 0.236) & (0.210, 0.234) & (0.207, 0.233) & (0.208, 0.234) & (0.209, 0.234) & \textbf{(0.207, 0.234)} & \textbf{(0.195, 0.223)} & \textbf{(0.180, 0.209)} & (0.313, 0.364) & (0.285, 0.297) & (0.244, 0.265) & (0.229, 0.255) & (0.459, 0.511) \\
1000 & 12 & 3 & 0.117 & 0.115 & 0.115 & \textbf{0.114} & \textbf{0.114} & 0.115 & \textbf{0.112} & \textbf{0.109} & 0.148 & 0.277 & 0.267 & 0.276 & 0.479 \\
\bottomrule
\end{tabular}
}
\end{table}

\end{landscape}

\begin{landscape}
  \begin{table}[ht]
\footnotesize
\caption{Cross-validated tdAUC (95\% bootstrap CI): common Weibull shape, normal errors.}
\label{nontruncated-tablesauctCommonShapeNormaltex}
\centering
\adjustbox{max width=\linewidth}{
\begin{tabular}[t]{cccrrrrrrrrrrrrr}
\toprule
N & landmark & horizon & Static-only & LOCF & LME & LCMM2 & LCMM3 & LCMM4 & LCMM2cl & LCMM3cl & LCMM4cl & JLCM1 & JLCM2 & JLCM3 & JLCM4
\\
\midrule
1000 & 9 & 3 & 0.473 & 0.565 & 0.598 & 0.582 & 0.569 & \textbf{0.601} & \textbf{0.622} & \textbf{0.660} & 0.496 & 0.498 & \textbf{0.618} & 0.446 & 0.440 \\
 &  &  & (0.422, 0.531) & (0.513, 0.618) & (0.542, 0.651) & (0.528, 0.636) & (0.513, 0.619) & \textbf{(0.546, 0.657)} & \textbf{(0.567, 0.680)} & \textbf{(0.605, 0.718)} & (0.449, 0.545) & (0.445, 0.554) & \textbf{(0.561, 0.676)} & (0.398, 0.499) & (0.389, 0.489) \\
1000 & 9 & 6 & 0.483 & 0.584 & 0.620 & 0.601 & 0.585 & 0.616 & \textbf{0.629} & \textbf{0.666} & 0.500 & 0.506 & 0.614 & 0.443 & 0.435 \\
 &  &  & (0.434, 0.529) & (0.540, 0.629) & (0.576, 0.665) & (0.560, 0.647) & (0.536, 0.629) & (0.574, 0.660) & \textbf{(0.585, 0.677)} & \textbf{(0.618, 0.713)} & (0.458, 0.545) & (0.465, 0.551) & (0.568, 0.662) & (0.400, 0.488) & (0.386, 0.476) \\
1000 & 9 & 9 & 0.484 & 0.601 & 0.643 & 0.626 & 0.613 & \textbf{0.645} & 0.640 & \textbf{0.685} & 0.505 & 0.491 & 0.589 & 0.434 & 0.427 \\
 &  &  & (0.440, 0.527) & (0.560, 0.644) & (0.599, 0.682) & (0.586, 0.668) & (0.568, 0.655) & \textbf{(0.607, 0.685)} & (0.598, 0.684) & \textbf{(0.644, 0.726)} & (0.464, 0.549) & (0.450, 0.535) & (0.547, 0.633) & (0.395, 0.476) & (0.383, 0.467) \\
1000 & 12 & 3 & 0.449 & 0.626 & 0.638 & 0.633 & \textbf{0.644} & \textbf{0.641} & 0.624 & \textbf{0.646} & 0.595 & 0.516 & 0.623 & 0.606 & 0.583 \\
\bottomrule
\end{tabular}
}
\end{table}

\end{landscape}

\begin{landscape}
  \begin{table}[ht]
\footnotesize
\caption{Cross-validated Brier score (95\% bootstrap CI): common Weibull shape, normal errors.}
\label{nontruncated-tablesbrierCommonShapeNormaltex}
\centering
\adjustbox{max width=\linewidth}{
\begin{tabular}[t]{cccrrrrrrrrrrrrr}
\toprule
N & landmark & horizon & Static-only & LOCF & LME & LCMM2 & LCMM3 & LCMM4 & LCMM2cl & LCMM3cl & LCMM4cl & JLCM1 & JLCM2 & JLCM3 & JLCM4
\\
\midrule
1000 & 9 & 3 & 0.137 & 0.136 & 0.135 & 0.136 & 0.136 & \textbf{0.135} & \textbf{0.133} & \textbf{0.129} & 0.224 & 0.242 & 0.238 & 0.419 & 0.475 \\
 &  &  & (0.120, 0.156) & (0.117, 0.155) & (0.117, 0.154) & (0.115, 0.155) & (0.117, 0.155) & \textbf{(0.118, 0.153)} & \textbf{(0.115, 0.151)} & \textbf{(0.112, 0.147)} & (0.207, 0.243) & (0.240, 0.243) & (0.226, 0.250) & (0.398, 0.439) & (0.453, 0.499) \\
1000 & 9 & 6 & 0.207 & 0.203 & 0.202 & 0.203 & 0.203 & \textbf{0.202} & \textbf{0.197} & \textbf{0.189} & 0.337 & 0.288 & 0.299 & 0.428 & 0.470 \\
 &  &  & (0.194, 0.221) & (0.188, 0.217) & (0.187, 0.217) & (0.187, 0.218) & (0.189, 0.218) & \textbf{(0.187, 0.216)} & \textbf{(0.182, 0.211)} & \textbf{(0.173, 0.206)} & (0.314, 0.360) & (0.282, 0.293) & (0.285, 0.314) & (0.406, 0.449) & (0.446, 0.495) \\
1000 & 9 & 9 & 0.240 & 0.232 & 0.230 & 0.232 & 0.232 & \textbf{0.229} & \textbf{0.222} & \textbf{0.209} & 0.358 & 0.306 & 0.340 & 0.423 & 0.451 \\
 &  &  & (0.232, 0.247) & (0.223, 0.241) & (0.220, 0.241) & (0.222, 0.242) & (0.223, 0.241) & \textbf{(0.220, 0.239)} & \textbf{(0.209, 0.234)} & \textbf{(0.195, 0.223)} & (0.333, 0.385) & (0.296, 0.317) & (0.319, 0.361) & (0.399, 0.446) & (0.425, 0.479) \\
1000 & 12 & 3 & 0.131 & 0.128 & 0.127 & 0.127 & \textbf{0.127} & \textbf{0.127} & \textbf{0.126} & \textbf{0.124} & 0.179 & 0.326 & 0.403 & 0.404 & 0.447 \\
\bottomrule
\end{tabular}
}
\end{table}

\end{landscape}
\subsubsection{Running times: simulation study}
\begin{table}[ht]
\centering
\caption{Summary of running times (hours) for every model considered in the simulation study of Section \ref{sec:simulation} ($N=2000$). For a fixed number of clusters, the running times of the LCMM-based landmarking pipelines is consistently lower than those of the JLCM. }
\label{tab:running_times_N2000}
\begin{tabular}{lrrrrrr}
\hline
Model                            &   Min      &      Q1       &    Median  &             Mean    &  Q3     &  Max     \\
\hline
Static-only                      &   0.001    &      0.001    &    0.001   &             0.001   &  0.001  &  0.001   \\
LOCF                             &   0.001    &      0.002    &    0.002   &             0.002   &  0.002  &  0.002   \\
LME                              &   0.002    &      0.002    &    0.002   &             0.002   &  0.002  &  0.002   \\
LCMM2                            &   0.107    &      0.125    &    0.132   &             0.138   &  0.139  &  0.305   \\
LCMM2cl                          &   0.108    &      0.123    &    0.133   &             0.139   &  0.140  &  0.300   \\
LCMM3                            &   0.488    &      0.611    &    1.706   &             7.398   &  7.649  &  47.105  \\
LCMM3cl                          &   0.486    &      0.603    &    0.824   &             5.319   &  6.309  &  32.345  \\
LCMM4                            &   2.247    &      2.857    &    3.833   &             7.259   &  7.147  &  28.252  \\
JLCM1                            &   0.003    &      0.003    &    0.003   &             0.003   &  0.004  &  0.005   \\
JLCM2                            &   0.248    &      0.300    &    0.338   &             0.342   &  0.376  &  0.458   \\
JLCM3                            &   1.201    &      1.410    &    1.520   &             1.921   &  1.764  &  4.945   \\
JLCM4                            &   3.834    &      5.339    &    6.498   &             10.461  &  8.447  &  34.809  \\
\hline
\end{tabular}
\end{table}
\subsection{Supplementary Figures}

\begin{figure}[t]
\centering\includegraphics[width=\textwidth]{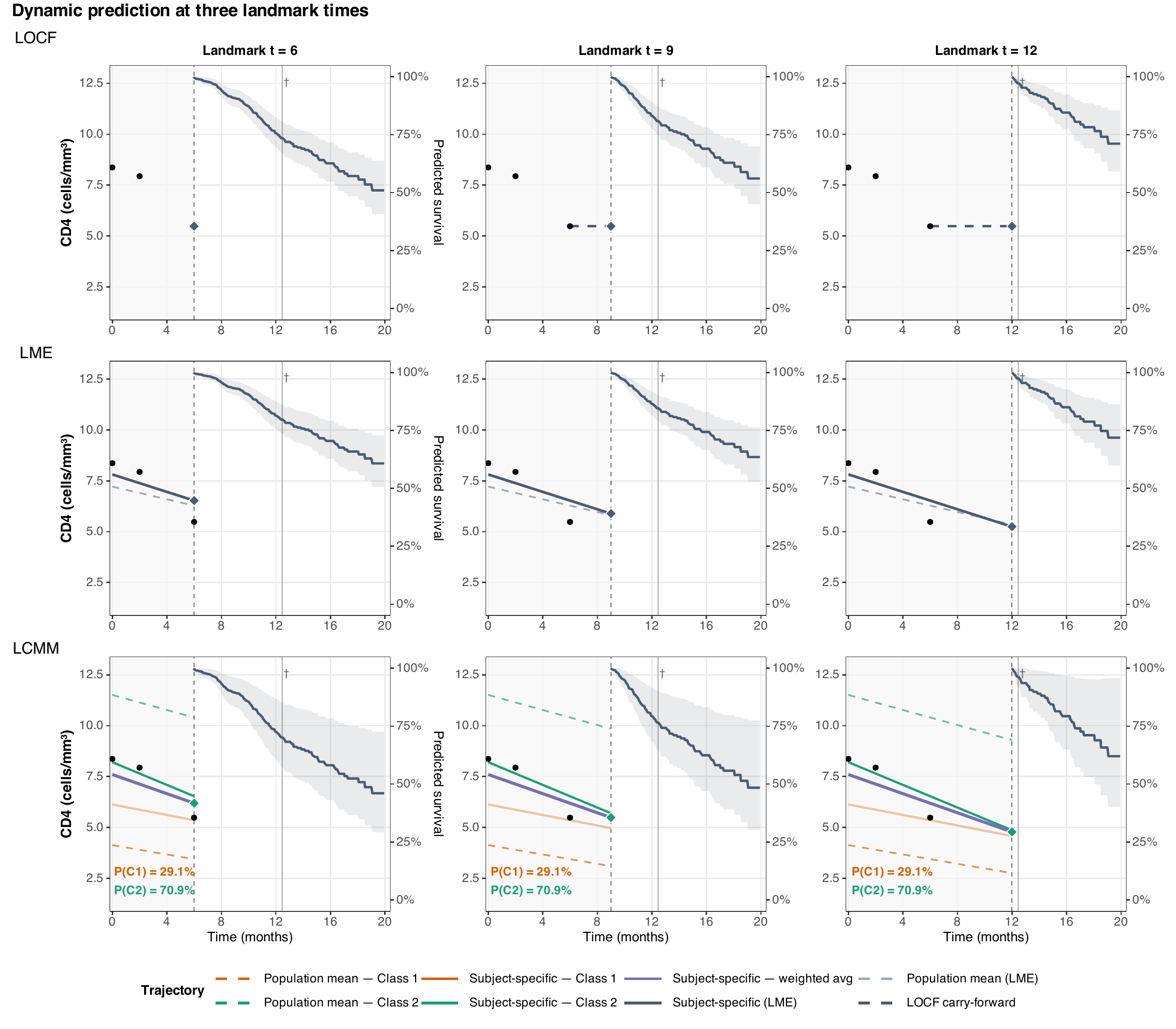}
        \caption{Illustration of dynamic prediction via landmarking using Last Observation Carried Forward (LOCF; top), Linear Mixed-Effect Models (LME; middle) and Latent Class Mixed Models (LCMM; bottom). At landmark times $s=6$ (left), $9$ (middle) and $12$ (right), landmarking dynamic prediction is performed, informed by a summary of the longitudinal biomarker measurements over $[0, s]$. }
        \label{fig:illustration}
\end{figure}
\begin{figure}[H]
    \centering
    \includegraphics[width=\textwidth]{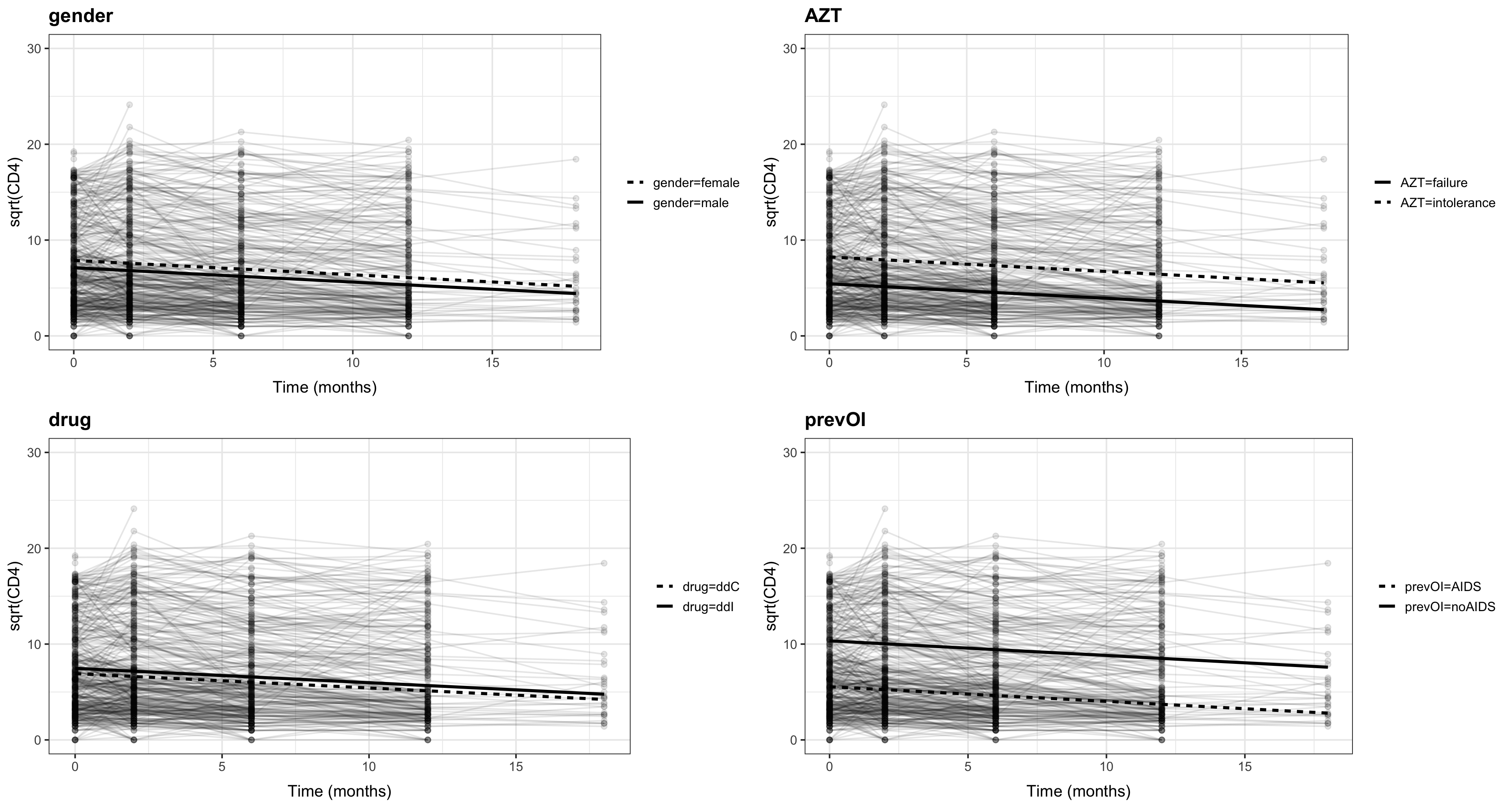}
    \caption{Longitudinal trajectories of sqrt-transformed CD4 counts in the \texttt{aids} for LMEs with a fixed effect for a single covariate: \texttt{gender} (top left), \texttt{AZT} (top right), \texttt{drug} (bottom left) and \texttt{prevOI} (bottom right). The dashed and solid lines are the estimated mean trajectories for each value of the binary covariate.}
    \label{fig:longitudinal-trajectories-alternative}
\end{figure}

\begin{figure}[htbp]
\centering
\includegraphics[width=\textwidth]{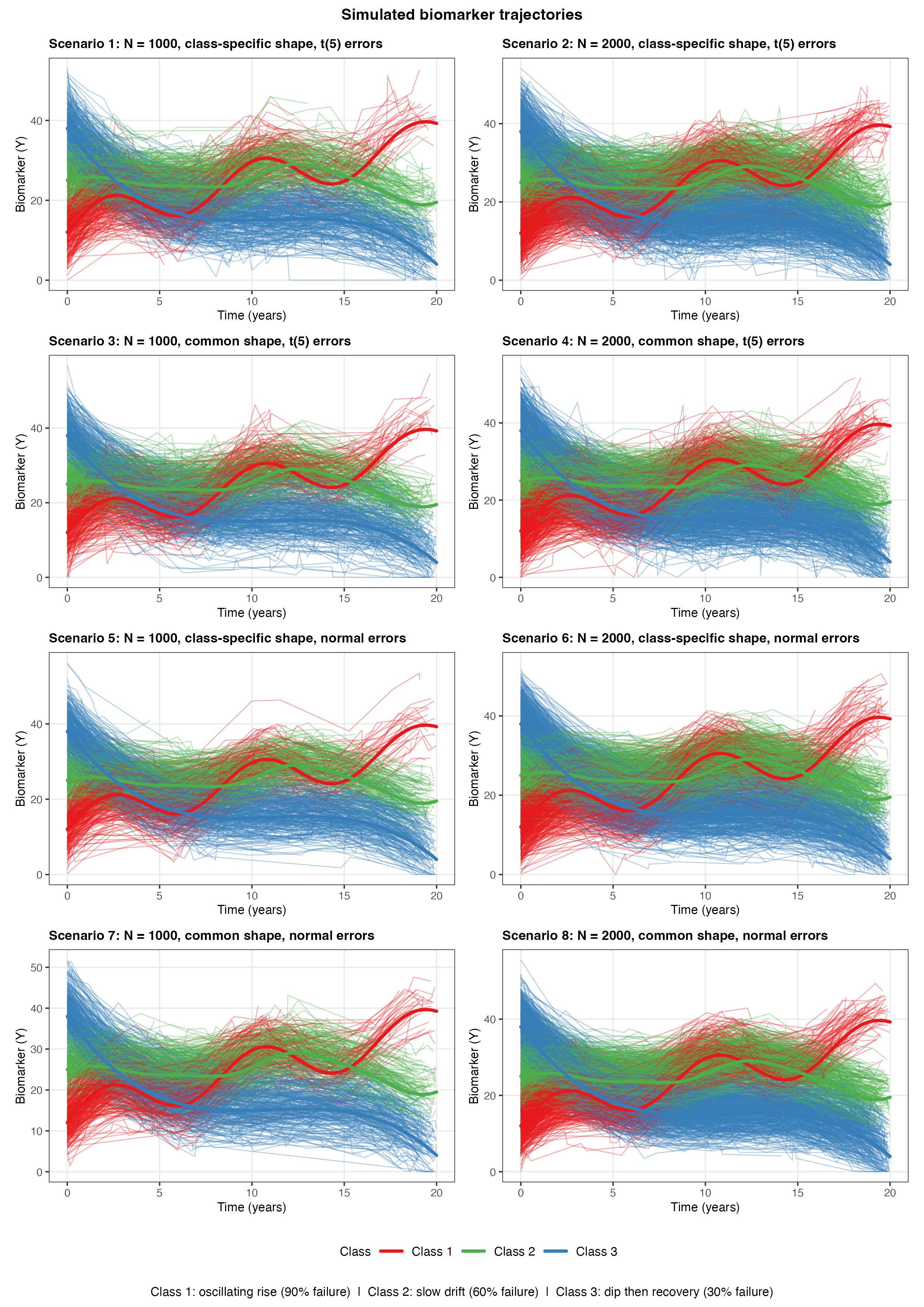}
\caption{Simulated biomarker trajectories across the eight scenarios. Thin lines represent individual trajectories coloured by true latent class, and thick lines overlay the ground-truth class-mean trajectories. Scenarios vary by sample size ($N \in \{1000, 2000\}$), shape specification (common vs.\ class-specific), and error distribution (normal vs.\ $t_5$).}
\label{fig:spaghetti-scenarios}
\end{figure}

\begin{figure}[ht]
    \centering
    \includegraphics[width=\textwidth]{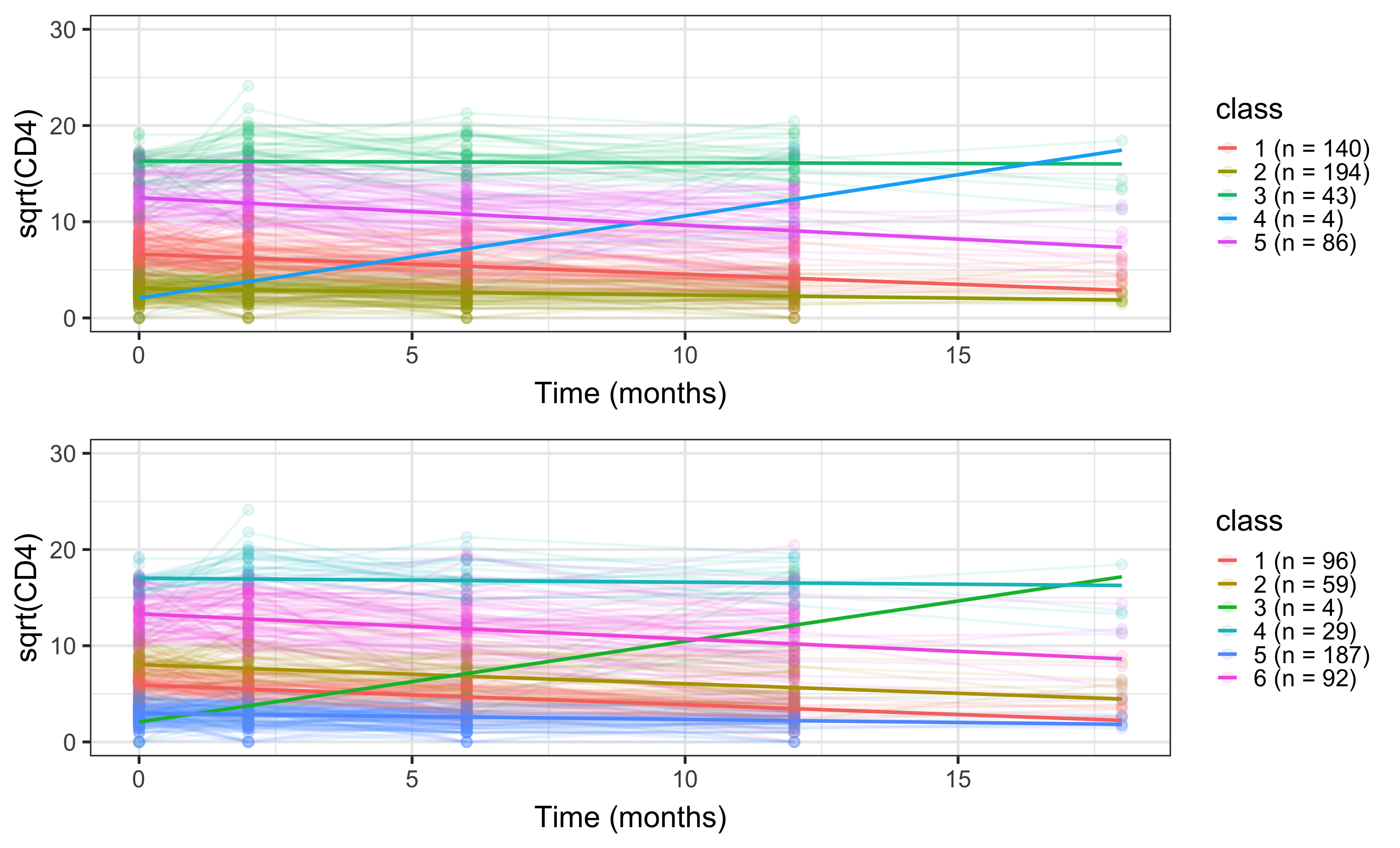}
    \caption{Longitudinal trajectories of square root-transformed CD4 counts in the \texttt{aids} dataset along with cluster membership allocation and estimated mean trajectories, according to LCMM with 5 (top) and 6 (bottom) clusters. The colours indicate cluster membership. Results indicate the presence of a small cluster of four individuals. }
    \label{fig:lcmm5and6}
\end{figure}

\begin{figure}[h]
    \centering
    \includegraphics[width=\textwidth]{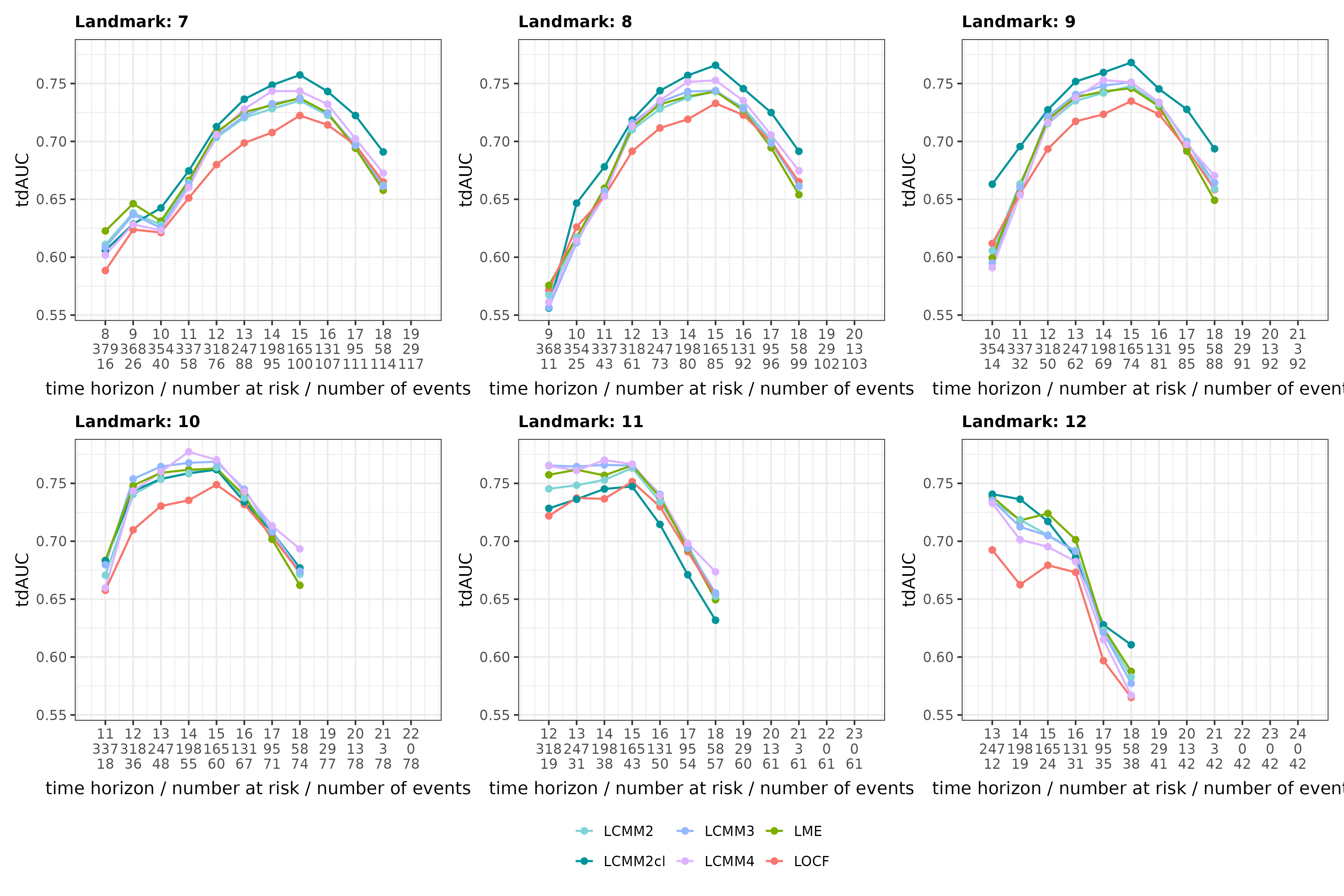}
    \caption{Cross-validated tdAUC for the different model specifications in the landmarking framework for the \texttt{aids} case study. For each specification, the increment in tdAUC is computed on the pooled out-of-sample predictions across the five cross-validation folds.}
    \label{fig:auct-performance-absolute}
\end{figure}

\begin{figure}[h]
    \centering
    \includegraphics[width=\textwidth]{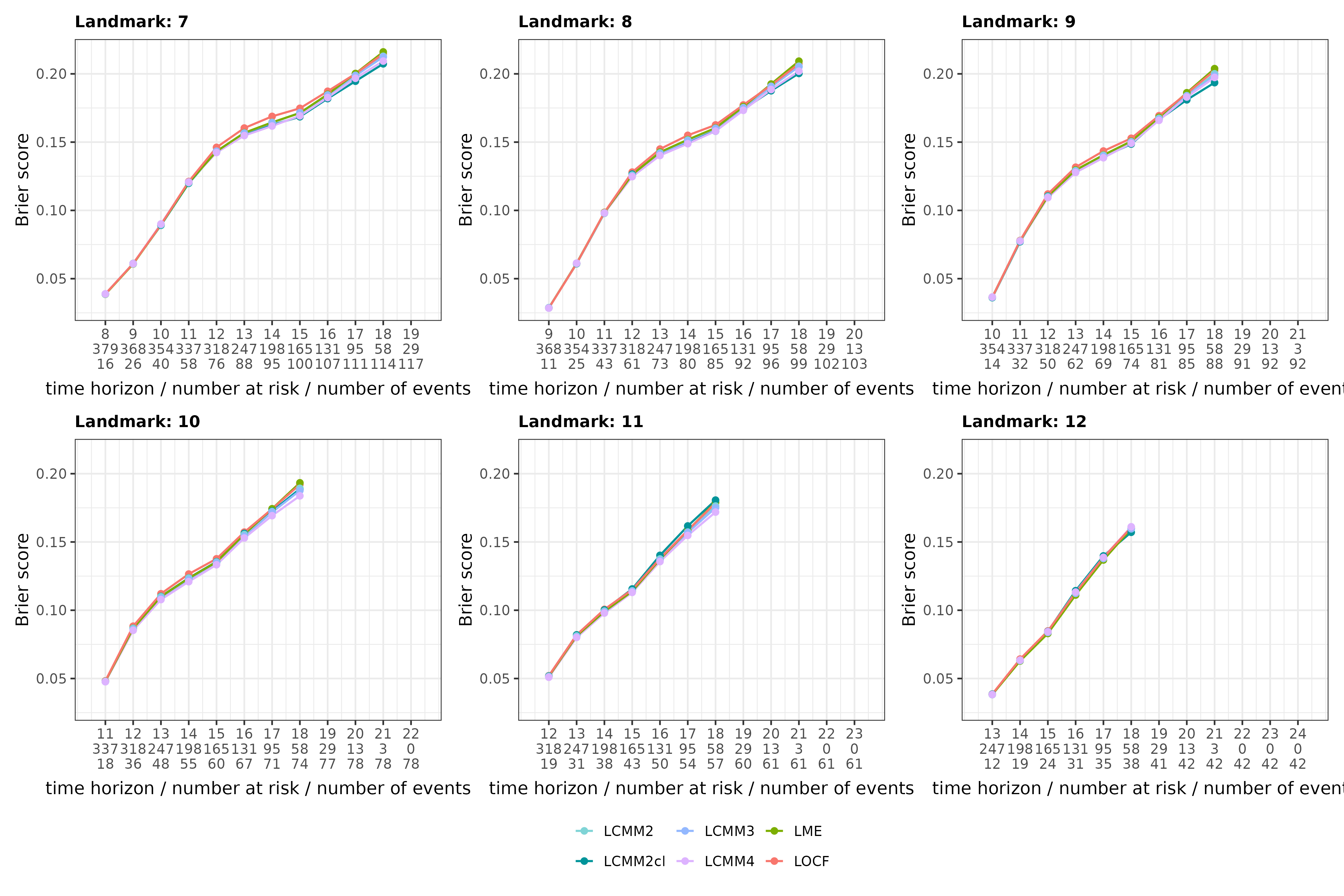}
    \caption{Cross-validated BS for the different model specifications in the landmarking framework for the \texttt{aids} case study. For each specification, the increment in tdAUC is computed on the pooled out-of-sample predictions across the five cross-validation folds. }
    \label{fig:brier-performance-absolute}
\end{figure}

\end{document}